\documentclass[pdflatex,sn-mathphys-num]{sn-jnl}

\usepackage{mathrsfs}%
\DeclareMathAlphabet{\mathscr}{OMS}{cmsy}{m}{n}
\usepackage{graphicx}%
\usepackage{multirow}%
\usepackage{amsmath,amssymb,amsfonts}%
\usepackage{amsthm}%

\usepackage[title]{appendix}%
\usepackage{xcolor}%
\usepackage{textcomp}%
\usepackage{manyfoot}%
\usepackage{booktabs}%
\usepackage{bm}%
\usepackage[algo2e]{algorithm2e}%
\usepackage{algorithm}%
\usepackage{algorithmicx}%
\usepackage{algpseudocode}%
\usepackage{listings}%
\usepackage{longtable}
\usepackage{svg}%
\usepackage[numbers]{natbib}%

\theoremstyle{thmstyleone}%
\newtheorem{theorem}{Theorem}
\newtheorem{proposition}[theorem]{Proposition}%

\theoremstyle{thmstyletwo}%

\theoremstyle{thmstylethree}%
\newtheorem{definition}{Definition}%

\newtheorem{corollary}{Corollary}
\begin{document}

\title[From  coherence to mixedness: a driver of barren plateaus in variational quantum algorithms]{From coherence to mixedness: a driver of barren plateaus in variational quantum algorithms}


\author*[1]{\fnm{Xiang} \sur{Zhou}}\email{zhoux2032@163.com}

\affil*[1]{\orgdiv{School of Mathematics and Computational Science}, \orgname{Shangrao Normal College}, \orgaddress{\city{Shangrao}, \postcode{334001}, \country{China}}}

\affil[2]{\orgdiv{School of Physical Science and Intelligent Education}, \orgname{Jiangxi Province Key Laboratory of Applied Optical Technology}, \orgaddress{\city{Shangrao}, \postcode{334001}, \country{China}}}


\abstract{
Variational quantum algorithms (VQAs) are a leading approach for near-term quantum advantage. However, their training is often hindered by barren plateaus (BPs). We present a framework based on observational entropy. The framework separates the coherent part of a quantum state from its incoherent part. We define the coherence fraction $\eta$ as the ratio of coherent to total contribution. This quantity measures how much of the coherence capacity is in a usable form. Using a 4-qubit Ising model and a hardware-efficient ansatz, we show that $\eta$ decreases monotonically during optimization, while the coherence capacity remains nearly constant. Our results show that $\eta$ provides an early indication of gradient collapse. It outperforms conventional diagnostics such as the gradient norm and entanglement entropy. This work offers a resource-theoretic explanation for BPs. It also provides a basis for real-time monitoring of coherence loss on noisy devices.
}




\maketitle

\section{Introduction}
\label{intro}
Variational quantum algorithms (VQAs) have become a widely used approach for near-term quantum computing \cite{Cerezo2021,Jiang2024}. They are a promising method for noisy intermediate-scale quantum (NISQ) devices \cite{RevModPhys.94.015004,Preskill2018quantumcomputingin,Cerezo2025,Fauseweh2024}. In practice, however, their performance is often limited by the barren plateau (BP) problem \cite{McClean2018,PRXQuantum.3.010313,Wang2021,Larocca2025,Sannia2024,PRXQuantum.5.030320,Dowling2026,OrtizMarrero2021}. In this problem, the gradient of the cost function decays exponentially with system size. As a result, training becomes ineffective because the landscape is nearly flat.

Recent research on the origins of BPs and possible mitigation strategies has advanced in several directions:

\begin{itemize}
    \item \textbf{Mechanisms}: BPs are related to circuit depth and entanglement. Noise can also induce BPs (NIBP)\cite{Wang2021,Schumann2024,Singkanipa2025,Qi2024}. Even a single noisy qubit can cause NIBP if the circuit is deep enough. This shows a general link between noise and BPs.
    
    \item \textbf{Structured models}: BPs can be avoided by using quantum models with problem-specific symmetries. West et al. designed rotation-equivariant models based on the quantum Fourier transform\cite{PRXQuantum.5.030320}. They used Lie algebra to prove that some submodels do not have BPs. These models performed better than general-purpose models in numerical tests. Restricting the search space to a relevant Hilbert subspace improves trainability.
    
    \item \textbf{Optimization strategies}: Boabang et al. (2026) proposed a two-stage least-squares method\cite{Boabang2026}. They use convex optimization for initialization to construct a smooth parameter basin. Then they switch to non-convex fine-tuning to avoid gradient-vanishing regions. The method performs well in cryptanalysis tasks.
    
    \item \textbf{Unconventional resources}: Some works explore using dissipation as a resource. Sannia et al. (2024) showed that model trainability can be preserved or restored by inserting Markovian dissipative channels after each circuit layer\cite{Sannia2024}. This strategy uses noise to combat noise.
    
    \item \textbf{Model architecture}: In generative quantum machine learning, Zhang et al. (2024) proposed the quantum denoising diffusion probabilistic model\cite{Zhang2026}. The model introduces intermediate training tasks between the target distribution and noise. This gradual structure helps avoid BPs while maintaining training efficiency and expressivity.
\end{itemize}

These studies have improved our understanding of BPs. However, a question remains. Why does optimization slow down even when entanglement does not decrease?

This observation shows that algorithm performance depends not only on the total amount of resources, but also on whether those resources are in a usable form. Standard resource theories quantify resources but do not distinguish between coherent and incoherent contributions \cite{PhysRevLett.113.140401,PhysRevLett.123.110402,PhysRevA.102.012411,PhysRevA.104.012404,PhysRevA.93.032136}.

Standard diagnostics such as the gradient norm and entanglement entropy provide a partial view of performance. This limitation is discussed in \cite{Wang2021}. The gradient norm often decreases only after optimization has slowed. Thus, it is not a good predictor of future performance. Entanglement entropy is sensitive to quantum correlations, but it can remain stable or even increase as performance declines. This makes it difficult to distinguish between coherence and mixedness.

Existing diagnostics do not track how resources change over time. We propose new diagnostic tools. These tools should:
\begin{itemize}
    \item Give early warnings of optimization failures.
    \item Distinguish between loss of coherence and conversion into mixedness.
    \item Show how coherence is converted into intra-block mixedness.
    \item Help stabilize training and avoid untrainable regimes.
\end{itemize}

We address this with a framework based on observational entropy (OE) \cite{PhysRevA.99.010101,PhysRevA.99.012103,Safranek2021,PhysRevA.102.052407,PRXQuantum.2.030202,Buscemi:2022wcd,Bai2024observational}. The framework decomposes the coherence capacity of a quantum state into two parts: coherent contributions and incoherent contributions. The coherent part captures inter-subspace superposition. The incoherent part, which we call intra-block mixedness, quantifies local classical disorder within each subspace. We define the coherence fraction $\eta$ as the ratio of the coherent part to the total coherence capacity. Under the conversion process, $\eta$ decreases as coherence is converted into mixedness, while the coherence capacity remains approximately conserved.

In this work, we focus on VQAs that use coherence for information processing. Incoherent contributions appear as noise within this framework. We treat an increase in intra-block mixedness as a negative indicator for coherence-driven algorithms. However, in some cases, intra-block mixedness can be useful. It may prevent overfitting to sharp coherent states and improve robustness. It may also help optimization escape narrow valleys and move to flatter regions. For tasks that prepare classical mixed states or solve classical optimization problems, intra-block mixedness is the target resource, not noise.

We test our framework on a 4-qubit transverse-field Ising model with $h=1$. We use a hardware-efficient ansatz with 32 parameters and the Adam optimizer ($\alpha_{\text{Adam}}=0.01$, $\beta_1=0.9$, $\beta_2=0.999$).

We find that during standard VQA optimization, coherent contributions are dynamic resources that gradients can use. In contrast, intra-block mixedness is a static or frozen resource.

The conversion of coherence into mixedness leads to optimization stagnation. Algorithm performance depends not only on the coherence capacity, but also on whether the resource form matches the algorithm's ability to use it.

Our framework connects quantum algorithm performance to resource conversion. It explains how optimization fails on NISQ devices. It also provides a way to diagnose and mitigate these failures through real-time monitoring of $\eta$.

The paper is organized as follows. Section~\ref{sec:2} introduces the necessary background on observational entropy and block coherence. Section~\ref{sec:3} presents the coherence-to-mixedness conversion mechanism and defines the coherence fraction $\eta$. Section~\ref{Sec:5} applies the framework to VQAs and shows how $\eta$ evolves during optimization. Section~\ref{sec:benchmarking} compares $\eta$ with standard diagnostics such as the gradient norm and entanglement entropy. Section~\ref{sec:3d_analysis} presents a three-dimensional analysis of $\eta$ across different conversion pathways. Section~\ref{con} summarizes the results and discusses future directions.

\section{Preliminaries}
\label{sec:2}

This section arms us with the essential foundations we need. These foundations enable us to truly master the theory developed in this work.
\subsection{Observational Entropy}
\label{sec:2.1}
Given a coarse-graining $\mathcal{C}=\{\hat{P}_{x}\}$ and an initial state $\rho$, the OE is defined as\cite{PhysRevA.99.010101}
\begin{align}\label{O01}
S_{\mathcal{C}}(\rho)=-\sum_{x}p_{x}\log_{2}\frac{p_{x}}{V_{x}},
\end{align}
where $p_{x}=\operatorname{Tr}[\hat{P}_{x}\rho\hat{P}_{x}]$ and $V_{x}=\operatorname{Tr}(\hat{P}_{x})$.

Following Refs.~\cite{Zhou2023,Buscemi:2022wcd,Bai2024observational}, we define
\begin{align}\label{O1}
\mathcal{O}_{\mathcal{C}}(\rho)=S_{\mathcal{C}}(\rho)-S(\rho),
\end{align}
where $S(\rho)=-\operatorname{Tr}(\rho\log_{2}\rho)$ is the von Neumann entropy. 

In our framework, we will refer to $\mathcal{O}_{\mathcal{C}}(\rho)$ as the coherence capacity of the state $\rho$ with respect to the coarse-graining $\mathcal{C}$.

This difference captures how much genuine quantum coherence the state retains. It provides a precise, quantitative measure of coherence. It obeys the following\cite{Zhou2023}:

(C1) $\mathcal{O}_{\mathcal{C}}(\rho)\geq 0$. 

(C2) $\mathcal{O}_{\mathcal{C}}(\rho)=0$ if and only if $\rho$ is exactly of the coarse-grained form $\rho=\sum_{x}\dfrac{p_{x}}{V_{x}}\hat{P}_{x}$. 

(C3) $\mathcal{O}_{\mathcal{C}}$ is invariant under any unitary that respects the coarse-graining: for every block-diagonal unitary $U=\bigoplus_{j}U_{j}$ commuting with $\mathcal{C}$,  
\begin{align}
\mathcal{O}_{\mathcal{C}}(U\rho U^{\dagger})=\mathcal{O}_{\mathcal{C}}(\rho).
\end{align}  

(C4) $\mathcal{O}_{\mathcal{C}}(\rho)$ is a convex functional of $\rho$, i.e., it never increases under probabilistic mixing of states.

These properties establish $\mathcal{O}_{\mathcal{C}}(\rho)$ as a powerful, coarse-graining–dependent quantifier of quantum coherence. It remains physically meaningful and versatile.

\subsection{Resource Theory of Inter-block Coherence}
\label{sec:2.2}
The resource theory of block coherence is originally proposed by {\AA}berg\cite{aberg2006quantifyingsuperposition}. It extends coherence theory from bases to general projective measurements. Let $\mathcal{H}=\bigoplus_{x}\mathcal{H}_{x}$ be a direct sum of orthogonal subspaces $\mathcal{H}_{x}$ with corresponding projectors $\hat{P}_{x}$. The set of block-incoherent (BI) states is
\begin{align}
\mathcal{I}_{\text{B}}(\mathcal{H})=\left\{\sum_{x}\hat{P}_{x}\sigma\hat{P}_{x} \,\middle|\, \sigma\in\mathcal{S}\right\}.
\end{align}

The corresponding block-dephasing operation is
\begin{align}
\Delta(\sigma)=\sum\limits_{x}\hat{P}_{x}\sigma\hat{P}_{x}.
\end{align}

Maximally block-incoherent (MBI) operations $\Lambda_{\mathrm{MBI}}$ are quantum channels that map BI states to BI states\cite{Marvian_2013,Marvian2014,PhysRevA.80.012307,PhysRevA.93.052331,PhysRevA.94.052324}:
\begin{align}
\Lambda_{\mathrm{MBI}}[\mathcal{I}_{\text{B}}(\mathcal{H})]\subseteq \mathcal{I}_{\text{B}}(\mathcal{H}).
\end{align}

A standard measure of block coherence is the relative-entropy measure
\begin{align}\label{O2}
\mathcal{C}_{\text{rel}}(\rho, \mathcal{C}) = S(\Delta(\rho)) - S(\rho).
\end{align}

It satisfies the following properties:
\begin{itemize}
    \item \textbf{Faithfulness:} $\mathcal{C}_{\text{rel}}(\rho, \mathcal{C}) \geq 0$, with equality if and only if $\rho$ is block-incoherent.
    \item \textbf{Monotonicity:} $\mathcal{C}_{\text{rel}}(\Lambda_{\text{MBI}}[\rho], \mathcal{C}) \leq \mathcal{C}_{\text{rel}}(\rho, \mathcal{C})$.
    \item \textbf{Convexity:} $\mathcal{C}_{\text{rel}}(\sum_x p_x \rho_x, \mathcal{C}) \leq \sum_x p_x \, \mathcal{C}_{\text{rel}}(\rho_x, \mathcal{C})$.
\end{itemize}

From a physical perspective, $\mathcal{C}_{\text{rel}}$ quantifies the additional uncertainty. This uncertainty is associated with coherent superpositions between different blocks. Block dephasing removes this inter-block coherence. It converts quantum superposition into classical mixtures and increases the von Neumann entropy.

\section{Coherence-to-Mixedness Conversion Mechanism}
\label{sec:3}
We now introduce the central quantities of our framework. They are defined in terms of the observational entropy and block coherence introduced in the previous section.

\subsection{Coherence Capacity}
For a coarse-graining $\mathcal{C}$ and state $\rho$, we have the identity\cite{PRXQuantum.2.030202}
\begin{align}
S_{\mathcal{C}}(\rho)=S\bigg[\sum_{x}\hat{P}_{x}\rho\hat{P}_{x}\bigg]+\sum_{x}p_{x}D[\rho_{x} \Vert \kappa_{x}],
\end{align}
where $\rho_x = \hat{P}_x \rho \hat{P}_x / p_x$ and $\kappa_x = \hat{P}_x / V_x$ are the maximally mixed states in block $x$. The quantity $D(\rho\Vert\sigma)=\mathrm{Tr}[\rho(\log_{2}\rho - \log_{2}\sigma)]$ is the quantum relative entropy.

We call $\mathcal{O}_{\mathcal{C}}$ the coherence capacity. Using the identity above, we obtain the following decomposition:
\begin{align}\label{O3}
\mathcal{O}_{\mathcal{C}}(\rho)=\mathcal{C}_{\text{rel}}(\rho, \mathcal{C})+\mathcal{D}_{\text{rel}}(\rho),
\end{align}
where $\mathcal{D}_{\text{rel}}(\rho)=\sum_{x}p_{x}D[\rho_{x} \Vert \kappa_{x}]$. We denote $\mathcal{C}_{\text{rel}}$ as inter-block coherence and $\mathcal{D}_{\text{rel}}$ as intra-block mixedness.

Equation~\eqref{O3} has a clear interpretation. The inter-block coherence $\mathcal{C}_{\text{rel}}$ captures quantum superposition between subspaces. The intra-block mixedness $\mathcal{D}_{\text{rel}}$ quantifies classical uncertainty within each subspace. Under free operations, $\mathcal{C}_{\text{rel}}$ can be converted into $\mathcal{D}_{\text{rel}}$.

\subsection{Coherence Fraction}
\label{sec:3.2}

\begin{definition}\label{Def1}
(Coherence Fraction) For a coarse-graining $\mathcal{C}=\{\hat{P}_{x}\}$ and an initial state $\rho$, the coherence fraction is defined as
\begin{align}
\eta(\rho):=\frac{\mathcal{C}_{\text{rel}}(\rho,\mathcal{C})}{\mathcal{O}_{\mathcal{C}}(\rho)},
\end{align}
with the convention that $\eta(\rho)=0$ when $\mathcal{O}_{\mathcal{C}}(\rho)=0$, since in that case, the state is fully consistent with the coarse-graining.
\end{definition}

\begin{itemize}
    \item \textbf{Physical meaning:} A value of $\eta$ close to 1 indicates that the coherence capacity is dominated by inter-block coherence, implying that the resource is in a form that is accessible for quantum information processing. In contrast, a value of $\eta$ close to 0 indicates that the coherence capacity is dominated by intra-block mixedness, meaning that the coherence has been largely converted into a less accessible form.
    \item \textbf{Properties:} By construction, $0 \leq \eta \leq 1$. Moreover, $\eta$ is non-increasing under free operations. For any free operation $\Lambda$, one has $\eta(\Lambda(\rho)) \leq \eta(\rho)$.
\end{itemize}

The definition of $\eta$ is motivated by the following observation: $\mathcal{O}_{\mathcal{C}}(\rho) = 0$ implies $\mathcal{C}_{\text{rel}}(\rho, \mathcal{C}) = 0$, but the converse does not hold\cite{Zhou2023}. Thus, $\eta$ measures the fraction of $\mathcal{O}_{\mathcal{C}}$ that comes from coherence. A large $\eta$ means coherence dominates. A small $\eta$ means intra-block mixedness dominates. In this sense, $\eta$ serves as a structure-sensitive metric. It enables comparisons between different resource states. It also guides optimization across various platforms.

\subsection{Coherence-to-mixedness conversion channel}
\label{sec:3.3}

The following definition precisely captures how resources transform under free operations. Ultimately, it lays the mathematical groundwork for building these channels.
\begin{definition}
(Resource Transformation Pathway) Consider a coarse-graining $\mathcal{C}=\{\hat{P}_{x}\}$ and an initial state $\rho$. Under a free operation $\Lambda$, the resource transformation pathway is given by
\begin{align}
\begin{aligned}
\Delta \mathcal{C}_{\text{rel}}&= \mathcal{C}_{\text{rel}}(\Lambda(\rho), \mathcal{C})  -  \mathcal{C}_{\text{rel}}(\rho, \mathcal{C}) \leq 0, \\
\Delta \mathcal{D}_{\text{rel}} &= \mathcal{D}_{\text{rel}}(\Lambda(\rho))  -  \mathcal{D}_{\text{rel}}(\rho), \\
\Delta \mathcal{O}_{\mathcal{C}} &= \mathcal{O}_{\mathcal{C}}(\Lambda(\rho))  -  \mathcal{O}_{\mathcal{C}}(\rho) = \Delta \mathcal{C}_{\text{rel}}+ \Delta \mathcal{D}_{\text{rel}},
\end{aligned}
\end{align}
where $\Delta \mathcal{C}_{\text{rel}}$ quantifies the decrease in $\mathcal{C}_{\text{rel}}$, and $\Delta\mathcal{D}_{\text{rel}}$ quantifies the change in $\mathcal{D}_{\text{rel}}$.
\end{definition}

Under this transformation pathway, we treat the following channels as free operations.

\begin{definition}\label{Def3}
(Coherence-to-mixedness conversion channel, CMCC) A conversion channel $\Lambda$ is a free operation that converts inter-block coherence into intra-block mixedness while approximately conserving the coherence capacity. A simple instance is

\begin{align}\label{Dc1}
\Lambda_{\alpha,\beta}(\rho) = (1-\alpha)\rho + \alpha\mathcal{D}_{\beta}(\rho),
\end{align}
where $\mathcal{D}_{\beta}(\rho)=\beta \Delta(\rho) + (1 - \beta) \sum\limits_x \hat{P}_x |\psi_0\rangle\langle\psi_0| \hat{P}_x$. Here $\alpha$ controls the conversion strength. The parameter $\beta$ sets the relative weight of the two conversion mechanisms.
\end{definition}

\begin{itemize}
  \item \textbf{Why is this operation free?}
  \begin{itemize}
    \item \textbf{$\Delta(\rho)$:} This map removes all inter-block coherence. It leaves only the block-diagonal part. Such dephasing is a standard free operation.
    \item \textbf{$\sum_x \hat{P}_x |\psi_0\rangle\langle\psi_0| \hat{P}_x$:} This map projects $|\psi_0\rangle$ into each block and forms a block-diagonal state. Preparing such states is free\cite{Marvian_2013}.
    \item \textbf{$\mathcal{D}_{\beta}(\rho)$:} This map combines the above two mechanisms with weight $\beta$. It is a free operation because each component is free.
    \item \textbf{$\Lambda_{\alpha,\beta}(\rho)$:} This channel is a convex mixture of the identity, a dephasing map, and a block-projection map. Since the set of free operations is convex, $\Lambda_{\alpha,\beta}$ is also free.
    \item \textbf{Role in this work:} The channel $\Lambda_{\alpha,\beta}$ converts coherence into intra-block mixedness in a controlled way. It allows us to study how $\eta$ changes during VQA optimization.
  \end{itemize}
\end{itemize}

To show the mathematical behavior and physical meaning of $\eta$, we now present and prove the following proposition.

\begin{proposition}\label{prop:eta_properties}
  The coherence fraction $\eta$ satisfies the following properties:
  \begin{enumerate}
    \item \textbf{Boundedness:} For any state $\rho$, $0 \leq \eta(\rho) \leq 1$.
    \item \textbf{Monotonicity under free operations:} For any free operation $\Lambda$ (i.e., any MBI operation), $\eta(\Lambda(\rho)) \leq \eta(\rho)$.
    \item \textbf{Strict decrease under CMCC:} For any CMCC $\Lambda_{\alpha,\beta}$ with $\alpha \geq 0$ and $\beta \in [0,1]$, we have $\eta(\Lambda_{\alpha,\beta}(\rho)) < \eta(\rho)$. Thus, the CMCC strictly decreases $\eta$. 
  \end{enumerate}
\end{proposition}

\textbf{Proof:} See Appendix~\ref{APPxA}.

The definition of CMCC immediately implies the following corollary.
\begin{corollary}\label{cor1}
 Under $\Lambda$, the change in $\eta$ is approximated by
    \begin{align}
       \eta( \Lambda(\rho)) \approx \eta(\rho) - \frac{|\Delta \mathcal{C}_{\text{rel}}|}{\mathcal{O}_\mathcal{C}(\rho)},
    \end{align}
where the ratio $\frac{|\Delta \mathcal{C}_{\text{rel}}|}{\mathcal{O}_\mathcal{C}(\rho)}$ measures the relative loss of coherence.
\end{corollary}

If $\mathcal{O}_\mathcal{C}(\rho)=0$, then $|\Delta \mathcal{C}_{\text{rel}}|=0$ and $\eta(\rho)=0$, and consequently $\eta( \Lambda(\rho))=0$. 

To cleanly compare the change in $\eta$, we define a relative metric. Let $\eta_{\text{init}}$ be the initial coherence fraction, and $\eta_{\text{final}}$ be the fraction after a period of evolution. The relative change is:

\begin{align}
\Delta\eta_{\%} = \frac{\eta_{\text{init}} - \eta_{\text{final}}}{\eta_{\text{init}}} \times 100\%,
\end{align}
where $\Delta\eta = \eta_{\text{init}} - \eta_{\text{final}}$. 

This metric is scale-invariant. It remains meaningful even when $\eta_{\text{init}}$ is small.

Using $\Delta\eta_{\%}$, we categorize the change in $\eta$ into three levels:

\begin{itemize}
    \item \textbf{Small change} ($\Delta\eta_{\%} \in (0, 25\%)$): The coherence fraction decreases slightly. The algorithm still improves, but less efficiently.
    
    \item \textbf{Moderate change} ($\Delta\eta_{\%} \in [25\%, 75\%)$): A large part of the block coherence is lost. Algorithm performance drops notably. Training often slows down or stalls.
    
    \item \textbf{Large change} ($\Delta\eta_{\%} \geq 75\%$): Most of the block coherence is gone. The algorithm is effectively trapped in a BP. Further optimization brings little improvement.
\end{itemize}

The next theorem is central to our work. It shows how resources change under a broad class of free operations.

\begin{theorem}\label{Thm1}
(Coherence-to-mixedness conversion) There exists a CMCC $\Lambda$ and a state $\rho$ such that
\begin{enumerate}
    \item Block coherence decreases: $\Delta\mathcal{C}_{\text{rel}} \leq 0$.
    \item Intra-block mixedness increases: $\Delta\mathcal{D}_{\text{rel}} \geq 0$.
    \item Coherence capacity is approximately conserved: $|\Delta\mathcal{D}_{\text{rel}}| \approx |\Delta\mathcal{C}_{\text{rel}}|$, so $\Delta\mathcal{O}_{\mathcal{C}} \approx 0$.
\end{enumerate}
Under these conditions, the coherence fraction $\eta$ decreases ($\Delta\eta \geq 0$). This shows a loss of coherence, while the coherence capacity remains nearly constant.
\end{theorem}

\textbf{Proof:} See Appendix~\ref{APPxB}. We provide an analytic example using a 4-dimensional system and a specific channel $\Lambda_\varepsilon$. For $\varepsilon = 0.26$, we obtain $\Delta\mathcal{C}_{\text{rel}} \approx -0.130$, $\Delta\mathcal{D}_{\text{rel}} \approx +0.129$, $\Delta\mathcal{O}_{\mathcal{C}} \approx -0.001$, and $\Delta\eta_{\%} \approx 49.376\%$, which confirms the mechanism.

This theorem shows that even if the coherence capacity is approximately conserved, the coherence fraction $\eta$ decreases. Coherence is converted into intra-block mixedness. As a result, the resource becomes less useful for the algorithm, even though the coherence capacity does not change.

\begin{figure}[ht]
\centering
\includegraphics[width=\linewidth]{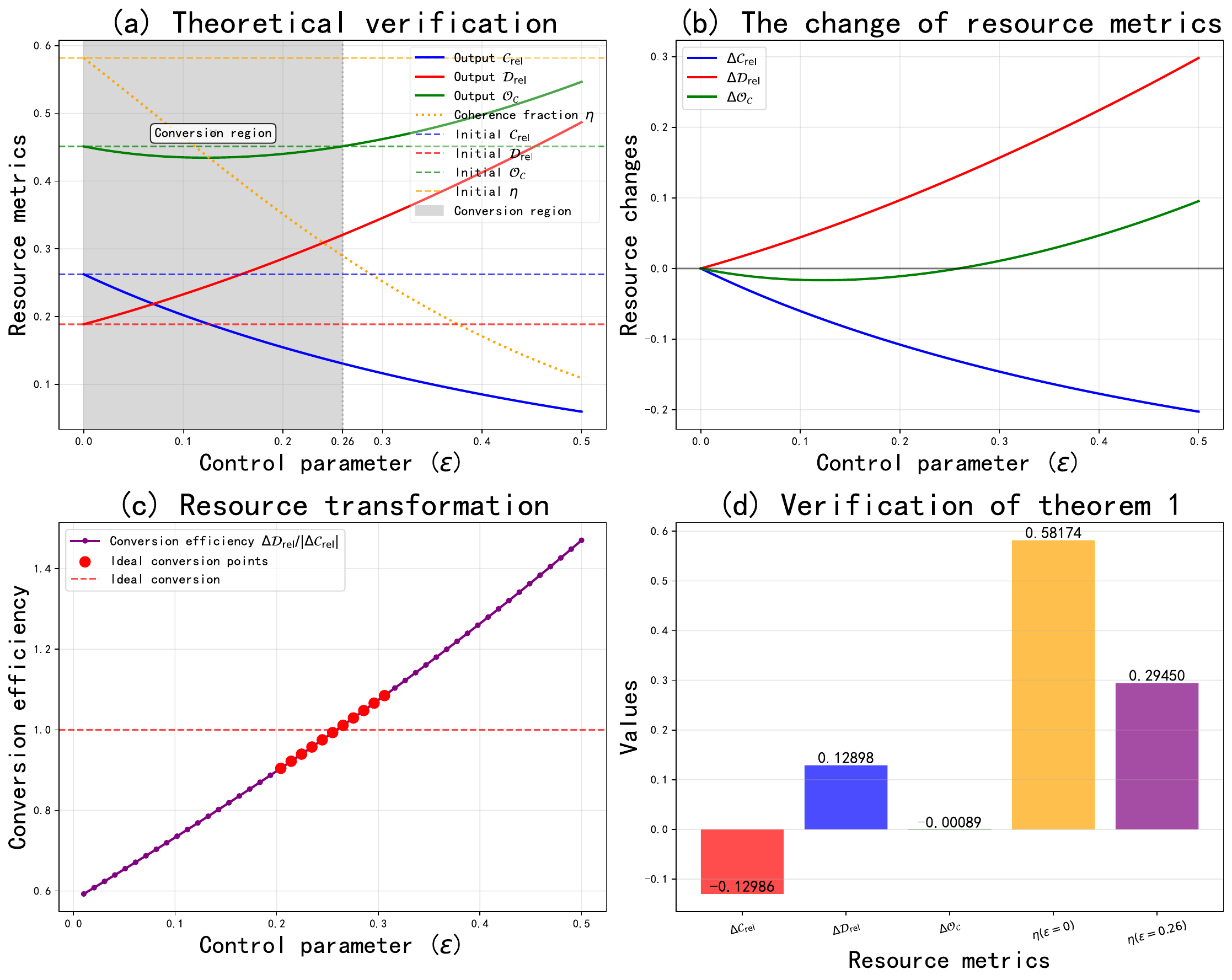}
\caption{\footnotesize{Coherence-to-mixedness conversion under the CMCC. (a) and (b): blue curves show $\mathcal{C}_{\text{rel}}$ and $\Delta\mathcal{C}_{\text{rel}}$, red curves show $\mathcal{D}_{\text{rel}}$ and $\Delta\mathcal{D}_{\text{rel}}$, green curves show $\mathcal{O}_{\mathcal{C}}$ and $\Delta\mathcal{O}_{\mathcal{C}}$, and the orange curve shows $\eta$. In (a), each quantity is plotted as a function of $\varepsilon$. In (b), the changes in these quantities are shown. (c) shows the ratio between $\Delta\mathcal{D}_{\text{rel}}$ and $|\Delta\mathcal{C}_{\text{rel}}|$. (d) shows that the three conditions of Theorem~\ref{Thm1} are satisfied at the critical point.}}
\label{figure1}
\end{figure}

Fig.~\ref{figure1} shows how the quantities in Theorem~\ref{Thm1} (see Appendix~\ref{APPxB}) evolve with $\varepsilon$.

Fig.~\ref{figure1}(a) tracks the metrics as functions of $\varepsilon$. The quantity $\mathcal{O}_{\mathcal{C}}$ shows a convex dependence on $\varepsilon$ over $[0,0.26]$. In contrast, $\eta$ decreases monotonically over this interval. It drops from about 0.58 to 0.29, a decrease of about $50\%$. This falls into the moderate change category defined above.

Fig.~\ref{figure1}(b) shows how each quantity responds to the channel. Here $\Delta\mathcal{C}_{\text{rel}}$ becomes negative, indicating a loss of block coherence. At the same time, $\Delta\mathcal{D}_{\text{rel}}$ becomes positive, indicating a gain in intra-block mixedness. Throughout this process, $\Delta\mathcal{O}_{\mathcal{C}}$ stays close to zero. This shows that coherence capacity is approximately conserved.

Fig.~\ref{figure1}(c) quantifies the conversion between $\mathcal{C}_{\text{rel}}$ and $\mathcal{D}_{\text{rel}}$. In some parameter regions, the ratio $\Delta\mathcal{D}_{\text{rel}}/|\Delta\mathcal{C}_{\text{rel}}|$ approaches one. This means that almost every unit of lost coherence becomes one unit of mixedness. This behavior agrees with our theoretical prediction.

Fig.~\ref{figure1}(d) checks the three conditions of Theorem~\ref{Thm1} at a critical point. All three hold: $\Delta\mathcal{C}_{\text{rel}}<0$, $\Delta\mathcal{D}_{\text{rel}}>0$, and $\Delta\mathcal{O}_{\mathcal{C}}\approx0$. The drop in $\eta$ indicates a loss of coherence, while the near-constant $\mathcal{O}_{\mathcal{C}}$ reinforces our main point.

\subsection{Summary and Discussion}

Our framework offers a way to diagnose quantum optimization. Instead of focusing on entanglement saturation or random parameters, we point to another mechanism. We identify the conversion of coherence into mixedness as a driver of optimization failure. At the center of this view is the coherence fraction $\eta$. 

\section{Coherence-to-Mixedness Conversion in VQAs}
\label{Sec:5}

In this section, we examine the mechanism behind optimization stagnation in VQAs. We link the slowdown of optimization, such as flat cost function plateaus, to the conversion of coherence into mixedness. This conversion is captured by the decrease of $\eta$. Our framework provides a single mechanism for the onset of BPs.

We apply our framework to VQAs and show that $\eta$ serves as a diagnostic for optimization stagnation. It signals the approach of BPs. Through controlled numerical experiments, we track how $\eta$ evolves during optimization. The results agree with the predictions of our theory.

\subsection{Experimental Setup}
\label{Sec:5.1}

\subsubsection{System Model and Hamiltonian}
\label{Sec:5.1.1}
We investigate a four-qubit transverse-field Ising model with periodic boundary conditions, $Z_{5}=Z_{1}$, described by the Hamiltonian\cite{Sachdev_2011}
\begin{align}
H = -\sum_{i=1}^{4} Z_{i} Z_{i+1} - h \sum_{i=1}^{4} X_{i} .
\end{align}

We set the transverse field to its quantum critical value, $h = 1$, which places the model at the critical point. This choice provides a clear basis for our coarse-graining scheme.

\subsubsection{Coarse-Graining Scheme}
\label{Sec:5.1.2}

We introduce projectors
\begin{align}
\hat{P}_{x} = \sum_{i \in I_{x}} \lvert E_{i} \rangle \langle E_{i} \rvert ,
\end{align}
where $I_{\tau}$ denotes the set of energy eigenstate indices belonging to the subspace $\tau$, and $|E_i\rangle$ are the energy eigenvectors obtained from diagonalizing the Hamiltonian $H$.

The volume (dimension) of each subspace is given by:
\begin{align}
V_{\tau} = \text{Tr}(\hat{P}_{\tau}) = |I_{\tau}|,
\end{align}
where $|I_{\tau}|$ represents the number of energy eigenstates in the subspace $\tau$.

For each subspace, we define the maximally mixed (uniform) reference state:
\begin{align}
\kappa_{\tau} = \frac{\hat{P}_{\tau}}{V_{\tau}} = \frac{1}{|I_{\tau}|} \sum_{i \in I_{\tau}} |E_i\rangle \langle E_i|.
\end{align}

These reference states serve as the baseline for quantifying intra-block mixedness through the quantum relative entropy $D(\rho_{\tau} \| \kappa_{\tau})$.

The Hilbert space is partitioned into two energy regions:
\begin{itemize}
    \item \textbf{Low-energy subspace}: $E < E_{\text{min}} + 0.5\Delta E$,
    \item \textbf{High-energy subspace}: $E \geq E_{\text{min}} + 0.5\Delta E$.
\end{itemize}
where $\Delta E = E_{\text{max}} - E_{\text{min}}$ represents the total energy range of the Hamiltonian spectrum. This 50\% energy threshold provides a balanced division of the energy eigenstates into two approximately equal-sized subspaces.

In the computational implementation, the projection operators are constructed as follows:

\begin{enumerate}
    \item Diagonalize the Hamiltonian to obtain eigenvalues $\{E_i\}$ and eigenvectors $\{|E_i\rangle\}$.
    \item Compute the energy range: $\Delta E = E_{\text{max}} - E_{\text{min}}$.
    \item Determine the threshold: $E_{\text{threshold}} = E_{\text{min}} + 0.5\Delta E$.
    \item Partition eigenstate indices:
    \begin{align}
        I_{\text{low}} &= \{i : E_i < E_{\text{threshold}}\}, \\
        I_{\text{high}} &= \{i : E_i \geq E_{\text{threshold}}\}.
    \end{align}
    \item Construct projection matrices:
    \begin{align}
        \hat{P}_{\text{low}} &= \sum_{i \in I_{\text{low}}} |E_i\rangle \langle E_i|, \\
        \hat{P}_{\text{high}} &= \sum_{i \in I_{\text{high}}} |E_i\rangle \langle E_i|.
    \end{align}
\end{enumerate}

This coarse-graining scheme has several advantages:
\begin{enumerate}
    \item \textbf{Clear structure}: With only two subspaces, the split of $\mathcal{O}_{\mathcal{C}}$ into $\mathcal{C}_{\text{rel}}$ and $\mathcal{D}_{\text{rel}}$ is more transparent and easier to interpret.
    
    \item \textbf{Coherence retention}: Fewer subspaces mean that less inter-block coherence $\mathcal{C}_{\text{rel}}$ is probed. This can slow the apparent decrease of $\eta$.
    
    \item \textbf{Balanced subspaces}: The $50\%$ threshold yields subspaces of comparable size. This avoids unbalanced coarse-grainings.
    
    \item \textbf{Computational efficiency}: Fewer projectors reduce the numerical cost of computing $\mathcal{O}_{\mathcal{C}}$ and its components.
\end{enumerate}

The projection operators satisfy the following properties:

\begin{enumerate}
    \item \textbf{Completeness}: $\hat{P}_{\text{low}} + \hat{P}_{\text{high}} = \mathbb{I}$.
    \item \textbf{Orthogonality}: $\hat{P}_{\tau} \hat{P}_{\tau'} = \delta_{\tau \tau'} \hat{P}_{\tau}$.
    \item \textbf{Idempotence}: $\hat{P}_{\tau}^2 = \hat{P}_{\tau}$.
    \item \textbf{Hermiticity}: $\hat{P}_{\tau}^{\dagger} = \hat{P}_{\tau}$.
\end{enumerate}

These properties ensure that the coarse-graining forms a valid projective measurement that respects the mathematical structure of quantum mechanics.

With this modified coarse-graining scheme, the key resource metrics are:

\begin{align}
    \mathcal{C}_{\text{rel}}(\rho, \mathcal{C}) &= S(\Delta(\rho)) - S(\rho), \\
    \mathcal{D}_{\text{rel}}(\rho) &= \sum_{\tau \in \{\text{low}, \text{high}\}} p_{\tau} D(\rho_{\tau} \| \kappa_{\tau}), \\
    \mathcal{O}_{\mathcal{C}}(\rho) &= \mathcal{C}_{\text{rel}}(\rho, \mathcal{C}) + \mathcal{D}_{\text{rel}}(\rho), \\
    \eta(\rho) &= \frac{\mathcal{C}_{\text{rel}}(\rho, \mathcal{C})}{\mathcal{O}_{\mathcal{C}}(\rho)},
\end{align}
where $p_{\tau} = \text{Tr}(\hat{P}_{\tau} \rho \hat{P}_{\tau})$ is the probability of finding the system in subspace $\tau$, and $\rho_{\tau} = \hat{P}_{\tau} \rho \hat{P}_{\tau} / p_{\tau}$ is the conditional state within that subspace.

\subsubsection{Cost Function}
\label{Sec:5.1.3}
In this subsection, we define the cost function for optimization in VQAs. 
\begin{align}
\mathcal{L}(\vec{\theta}) = \langle \psi(\vec{\theta}) | H | \psi(\vec{\theta}) \rangle,
\end{align}
where $|\psi(\vec{\theta})\rangle$ is the ansatz state prepared by the parameterized quantum circuit $U(\vec{\theta})$, and $\vec{\theta}$ denotes the set of tunable circuit parameters.

\subsubsection{Circuit Architecture}
\label{Sec:5.1.4}
We employ a hardware-efficient ansatz\cite{PRXQuantum.3.010313,Wang2021}
\begin{align}\label{U1}
U(\vec{\theta}) = \left( \prod_{l=1}^{4} \mathcal{L}_l(\vec{\theta}) \right) \left( \prod_{i=0}^{3} H_i \right),
\end{align}
where
\begin{align}
\mathcal{L}_l(\vec{\theta}) = \left( \prod\limits_{i=0}^{3} R_x(\vec{\theta}^{x}_{l,i}) R_y(\vec{\theta}^{y}_{l,i}) \right)
\left( \prod\limits_{i=0}^{3} CX_{i,i+1} \right),
\end{align}
represents the $l$th circuit layer.

The circuit prepares the ansatz state $|\psi(\vec{\theta})\rangle$ by first applying Hadamard gates to all qubits. Then it applies $L = 4$ identical layers. Each layer $\mathcal{L}_l(\vec{\theta})$ contains parameterized single-qubit rotations and a chain of CNOT gates. The CNOT gates generate entanglement. The circuit has 32 tunable parameters in total.

\subsubsection{Optimization Protocol}
\label{Sec:5.1.5}
In this subsection, we describe the three-phase conversion protocol. The protocol gradually drives the system from a clean optimization regime into a regime with strong conversion effects. The conversion strength is controlled by the parameter $\alpha$ of the CMCC.

\begin{itemize}
\item \textbf{Phase I (0--30):} No conversion ($\alpha = 0$). The optimization landscape is clean. The algorithm explores the full Hilbert space and finds the optimum. This phase provides a baseline for performance under ideal conditions.

\item \textbf{Phase II (30--60):} Linear increase of conversion strength ($\alpha$ from $0$ to $0.5$). This phase introduces mild to moderate conversion. It is similar to running a medium-depth circuit on NISQ devices. We track how the decrease of $\eta$ slows down optimization.

\item \textbf{Phase III (60--150):} Strong conversion ($\alpha$ from $0.5$ to $0.9$). In this regime, the conversion is strong. The coherence is largely lost. The parameters continue to update, but the cost function stops improving. The algorithm enters a BP. Further training brings little benefit.
\end{itemize}

\subsection{Adam Optimizer}
\label{Sec:5.2}
We use the standard Adam optimizer for parameter updates. Adam is an adaptive stochastic method. It uses first-moment estimation (momentum) and second-moment estimation for adaptive learning rates. These features make Adam suitable for high-dimensional non-convex optimization.
\subsubsection{Parameter Update Rule}
\label{Sec:5.2.1}
The standard Adam update rule is as follows. At the $k$-th iteration, the parameter update $\Delta \theta^{(k)}$ is computed via:
\begin{enumerate}
    \item \textbf{Compute the gradient}:
    \begin{align}
    g_k = \nabla_{\theta} \mathcal{L}(\theta^{(k)}).
    \end{align}
    
    \item \textbf{Update the first-moment estimate}:
    \begin{align}
    m_k = \beta_1 m_{k-1} + (1 - \beta_1) g_k.
    \end{align}
    where $\beta_1$ represents the exponential decay rate for the first moment.
    
    \item \textbf{Update the second-moment estimate}:
    \begin{align}
    v_k = \beta_2 v_{k-1} + (1 - \beta_2) g_k^2.
    \end{align}
     where $\beta_2$ represents the exponential decay rate for the second moment.
    \item \textbf{Bias correction}:
    \begin{align}
    \hat{m}_k = \frac{m_k}{1 - \beta_1^k}, \quad \hat{v}_k = \frac{v_k}{1 - \beta_2^k}.
    \end{align}
    
    \item \textbf{Parameter update}:
    \begin{align}
    \Delta \theta^{(k)} = - \alpha_{\text{Adam}} \cdot \frac{\hat{m}_k}{\sqrt{\hat{v}_k} + \epsilon},
    \end{align}
where $\alpha_{\text{Adam}}$ is the learning rate, and $\epsilon$ is the numerical stability constant. Let $\epsilon = 10^{-8}$.
\end{enumerate}

The full update is then:
\begin{align}
\theta^{(k+1)} = \theta^{(k)} + \Delta \theta^{(k)}.
\end{align}

\subsubsection{Parameter Configuration}
\label{Sec:5.2.2}
We use the following fixed hyperparameter values throughout the optimization:

\begin{align}
\alpha_{\text{Adam}} = 0.01, \quad \beta_1 = 0.9,\quad \beta_2 = 0.999.
\end{align}

The first-moment vector $m_0$ and the second-moment vector $v_0$ are initialized as zero vectors with the same dimension as the parameter vector $\theta$. At the same time, gradients $g_k$ are computed via the parameter-shift rule of the parameterized quantum circuit at each iteration.

\subsection{Coherence-to-Mixedness Conversion: Experimental Results}
\label{Sec:5.3}

We now present experimental results based on the setup described in Sec.~\ref{Sec:5.1} and Sec.~\ref{Sec:5.2}. By tracking the evolution of $\eta$, we show that the conversion of coherence into mixedness is a cause of optimization stagnation in VQAs.

The Pearson correlation coefficient $r$ measures the linear relationship between two variables. It is defined as\cite{10.1098/rspl.1895.0041,Qi2017}
\begin{align}
 r = \frac{\sum_{k=1}^{n}(x_k - \bar{x})(y_k - \bar{y})}{\sqrt{\sum_{k=1}^{n}(x_k - \bar{x})^2 \sum_{k=1}^{n}(y_k - \bar{y})^2}},
\end{align}
where $x_k = \mathcal{C}_{\text{rel}}$ and $y_k = \mathcal{D}_{\text{rel}}$ at iteration $k$. The quantities $\bar{x}$ and $\bar{y}$ are the sample means, and $n$ is the sample size.

The uncertainty of $r$ is obtained using the Fisher $Z$ transformation\cite{Fisher1992}. This transformation provides confidence intervals for correlation coefficients.

We apply the Fisher $Z$ transformation:
\begin{align}
 Z = \frac{1}{2} \log_{2} \left( \frac{1 + r}{1 - r} \right).
\end{align}

The standard error of $Z$ is
\begin{align}
 SE_Z = \frac{1}{\sqrt{n - 3}}.
\end{align}

For a 95\% confidence interval, the bounds in the $Z$ domain are
\begin{align}
 Z_{\text{lower}} &= Z - 1.96 \times SE_Z, \\
 Z_{\text{upper}} &= Z + 1.96 \times SE_Z.
\end{align}

Then, we transform these bounds back to the correlation domain. For a given $Z$, we have
\begin{align}
 r = \frac{e^{2Z} - 1}{e^{2Z} + 1},
\end{align}
which is applied to $Z_{\text{lower}}$ and $Z_{\text{upper}}$ to obtain the confidence interval for $r$.

This statistical characterization of the correlation between $\mathcal{C}_{\text{rel}}$ and $\mathcal{D}_{\text{rel}}$ provides a precise and quantitative analysis. It quantifies the degree to which these resource metrics co-vary across successive iterations. Moreover, it is formulated in a manner that is directly compatible with standard correlation methodologies. More importantly, these methodologies are widely employed in contemporary physics research.

\begin{figure}[ht]
\centering
\includegraphics[width=\linewidth]{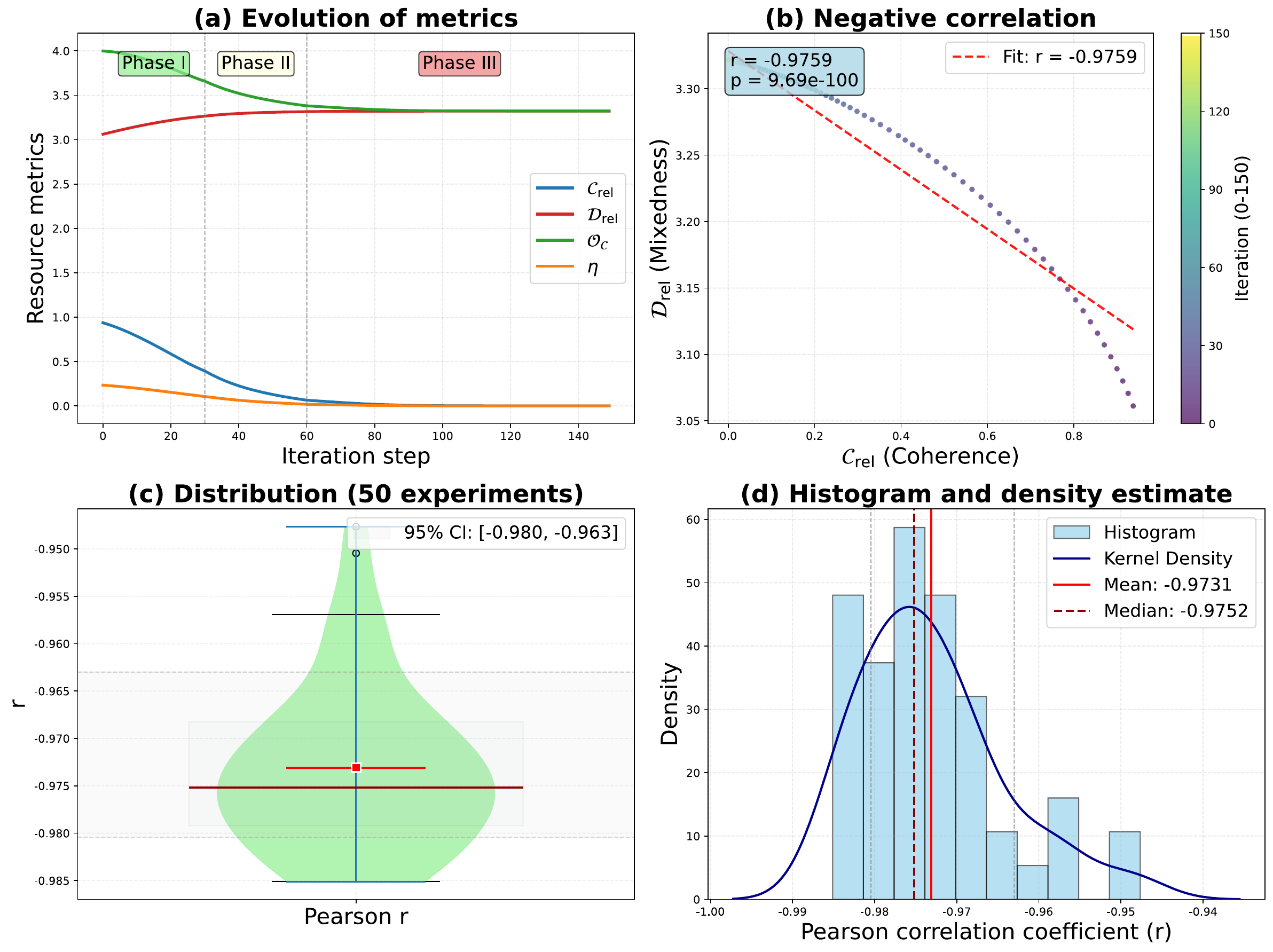}
\caption{\footnotesize{Coherence-to-mixedness conversion with two-subspace coarse-graining and the CMCC $\Lambda_{\alpha,\beta}$. (a): Evolution of the quantities across the three-phase protocol with $\beta=0.74$. The coherence $\mathcal{C}_{\text{rel}}$ (blue) decreases and the mixedness $\mathcal{D}_{\text{rel}}$ (red) increases. The coherence capacity $\mathcal{O}_{\mathcal{C}}$ (green) stays approximately constant. The coherence fraction $\eta$ (orange) decreases monotonically. (b): Correlation between $\mathcal{C}_{\text{rel}}$ and $\mathcal{D}_{\text{rel}}$ during the three phases. The linear fit gives $r = -0.9762$ and $p \approx 0$. (c): Distribution of $r$ from 50 independent experiments. The boxplot (light blue) shows quartiles. The violin plot (light green) shows density. The red square marks the mean ($r = -0.9731$). The 95\% confidence interval (gray band) is $[-0.980, -0.963]$. (d): Histogram and kernel density estimate of $r$. The distribution is left-skewed with a mean near $-0.9752$.}}
\label{figure2}
\end{figure}

Fig.~\ref{figure2} summarizes the results from Sec.~\ref{Sec:5.1} and Sec.~\ref{Sec:5.2}. It shows that $\mathcal{C}_{\text{rel}}$ is converted into $\mathcal{D}_{\text{rel}}$ during optimization. This conversion reduces $\eta$.

In Phase I (no conversion), $\mathcal{C}_{\text{rel}}$ and $\eta$ remain relatively high. In Phase II, $\mathcal{C}_{\text{rel}}$ decreases while $\mathcal{D}_{\text{rel}}$ increases. In Phase III, $\mathcal{C}_{\text{rel}}$ continues to decrease, $\mathcal{D}_{\text{rel}}$ continues to increase, and $\eta$ declines steadily.

The Adam optimizer uses momentum ($\beta_1 = 0.9$) and adaptive learning rates ($\beta_2 = 0.999$). Its update rule is
\begin{align}
\theta^{(k+1)} = \theta^{(k)} - \alpha_{\text{Adam}} \cdot \frac{\hat{m}_t}{\sqrt{\hat{v}_t} + \epsilon},
\end{align}
with $\alpha_{\text{Adam}} = 0.01$. As optimization proceeds, the parameters move toward low-energy regions. This reduces quantum coherence and steers the system toward classical-like states.

The parameter $\alpha$ increases linearly from 0 to 0.9. The CMCC $\Lambda_{\alpha,\beta}$ acts on the system at each iteration and accelerates the conversion.

With $\beta = 0.74$, $74\%$ of the conversion comes from block dephasing $\Delta(\rho)$, and $26\%$ comes from projection onto the ground state. This dephasing-dominated regime converts $\mathcal{C}_{\text{rel}}$ into $\mathcal{D}_{\text{rel}}$.

The observed trend shows $\mathcal{C}_{\text{rel}}$ approaching zero while $\mathcal{D}_{\text{rel}}$ grows. This has several consequences:

\begin{itemize}
    \item \textbf{Loss of block coherence}: When $\mathcal{C}_{\text{rel}} \approx 0$, the state becomes block-incoherent. Coherent superpositions between energy subspaces are lost. Only classical mixtures within each block remain.
    
    \item \textbf{Growth of intra-block mixedness}: The increase of $\mathcal{D}_{\text{rel}}$ shows that the lost coherence is converted into intra-block mixedness. Large $\mathcal{D}_{\text{rel}}$ values indicate that the state within each block deviates from the maximally mixed state.
    
    \item \textbf{Decrease of coherence fraction}: The ratio $\eta = \mathcal{C}_{\text{rel}} / \mathcal{O}_{\mathcal{C}}$ approaches zero. The denominator $\mathcal{O}_{\mathcal{C}} = \mathcal{C}_{\text{rel}} + \mathcal{D}_{\text{rel}}$ remains finite. The coherence capacity is conserved, but the coherence fraction decreases. This is the mechanism described in Theorem~\ref{Thm1}.
    
    \item \textbf{Effect of Adam}: As gradients become small in flat regions, Adam's updates shrink. The conversion channel then dominates the state evolution. This accelerates the loss of coherence.
\end{itemize}

The total decrease of $\eta$ exceeds $99\%$. This occurs because:

\begin{enumerate}
    \item With $\beta=0.74$, the CMCC $\Lambda_{\alpha,\beta}$ removes $\mathcal{C}_{\text{rel}}$ through dephasing.
    \item Adam steers parameters into regions where coherence offers little advantage for lowering the energy.
    \item Because $\mathcal{O}_{\mathcal{C}}$ is approximately conserved, the lost coherence is converted into $\mathcal{D}_{\text{rel}}$. This pushes $\eta$ toward zero.
\end{enumerate}

The correlation in Fig.~\ref{figure2}(b) gives $r = -0.9759$ and $p = 9.69 \times 10^{-100}$. This indicates a strong negative correlation between $\mathcal{C}_{\text{rel}}$ and $\mathcal{D}_{\text{rel}}$.

To test the reliability of this correlation, we perform 50 independent runs with different random seeds. Table~\ref{tab:correlation_stats} summarizes the results.

\begin{table}[htbp]
    \centering
    \caption{
        \footnotesize{Statistical summary of 50 independent runs at $\beta = 0.74$. The table shows the number of experiments, mean $r$, median $r$, standard deviation of $r$, 95\% confidence interval for the mean, mean $p$-value, $t$-statistic, and $p$-value of the $t$-test.}
    }
    \label{tab:correlation_stats}
    \setlength{\tabcolsep}{14pt}
    \begin{tabular}{lr}
        \hline
        \textbf{Statistic} & \textbf{Value} \\
        \hline
        \textbf{Number of experiments} & 50 \\
        \textbf{Mean Pearson $r$} & $-0.9731$ \\
        \textbf{Median Pearson $r$} & $-0.9752$ \\
        \textbf{Standard deviation of $r$} & $0.0087$ \\
        \textbf{95\% CI lower bound} & $-0.9804$ \\
        \textbf{95\% CI upper bound}& $-0.9630$ \\
        \textbf{Mean $p$-value} & $5.93 \times 10^{-77}$ \\
        \textbf{$t$-statistic (vs. zero)} & $-780.5919$ \\
        \textbf{$p$-value of $t$-test} & $5.43 \times 10^{-102}$ \\
        \hline
    \end{tabular}
\end{table}

As shown in Table~\ref{tab:correlation_stats}, the mean $r$ is $-0.9731$ with a standard deviation of $0.0087$. The 95\% confidence interval is $[-0.9804, -0.9630]$. All 50 runs give $p < 0.001$. The one-sample $t$-test gives $t = -780.5919$ and $p = 5.43 \times 10^{-102}$. These results confirm a strong and reproducible negative correlation between $\mathcal{C}_{\text{rel}}$ and $\mathcal{D}_{\text{rel}}$.

\subsection{Summary and Discussion}
The conversion channel $\Lambda_{\alpha,\beta}$ converts coherence into mixedness. This reduces $\eta$ and drives the system toward a BP. The results show that BPs in VQAs can arise from the conversion of coherence into mixedness, even when the coherence capacity appears to be conserved. The experimental protocol tracks this conversion and supports our theoretical framework.
\section{Comparison with Standard Diagnostics}
\label{sec:benchmarking}

In this section, we compare the coherence fraction $\eta$ with two standard diagnostics: the gradient norm and entanglement entropy. The comparison uses the VQA setup described in Sec.~\ref{Sec:5}. We test how well $\eta$ serves as an early-warning signal.

\subsection{Coherence fraction as a Diagnostic: Early Warning and Predictive Analysis}

The results in this subsection come from a single optimization run with $\beta = 0.74$ and the full three-phase protocol.

\begin{figure}[ht]
\centering
\includegraphics[width=\linewidth]{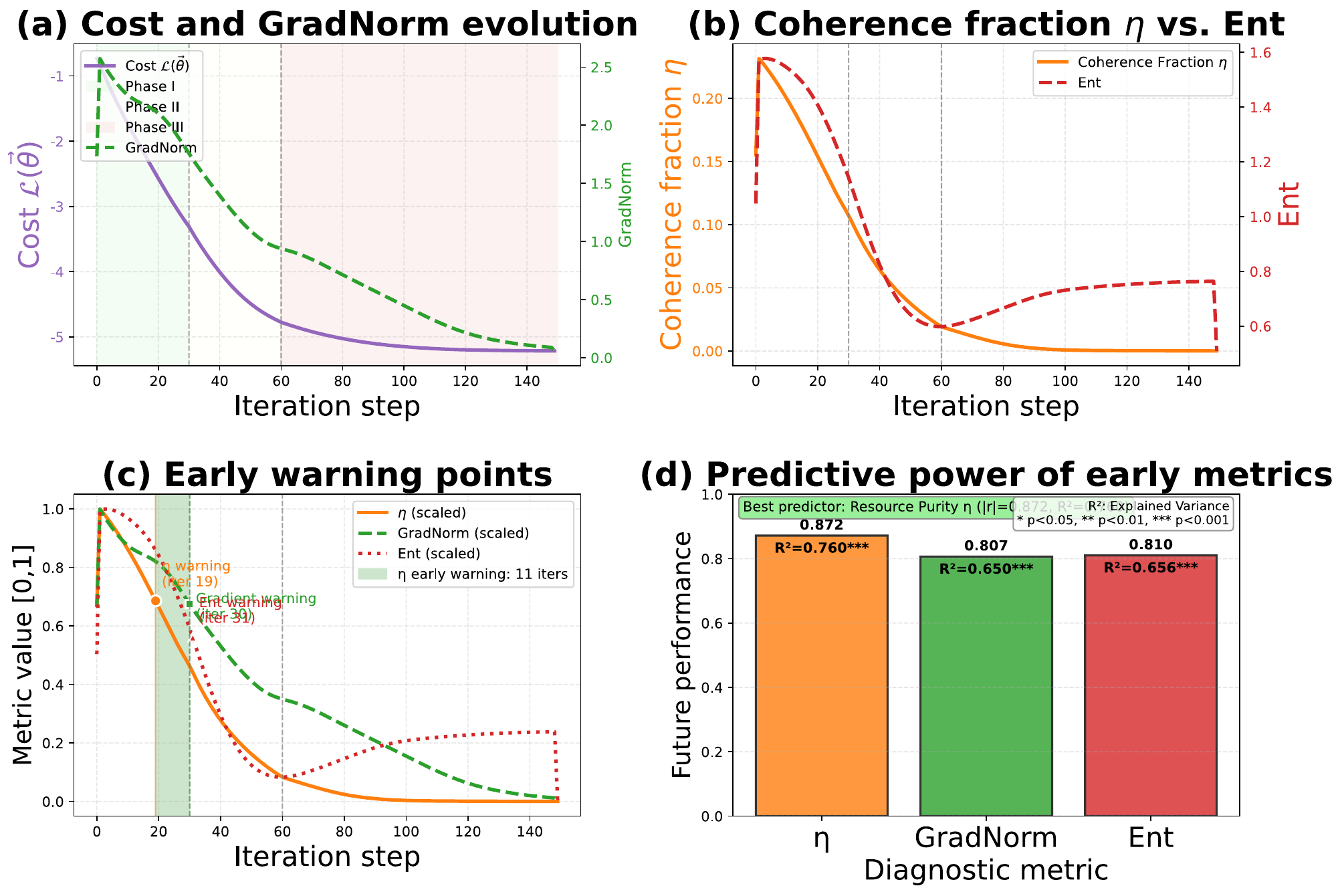}
\caption{\footnotesize{Comparison of $\eta$ with standard diagnostics under the three-phase protocol with $\beta = 0.74$. (a) Cost function (purple solid) and gradient norm (green dashed). (b) Coherence fraction $\eta$ (orange solid) and entanglement entropy (red dashed). (c) Iteration at which each metric reaches selected decline thresholds. (d) Correlation of each metric with future performance gains.}}
\label{figure3}
\end{figure}

Fig.~\ref{figure3}(a) shows the cost function $\mathcal{L}(\vec{\theta})$ and the gradient norm over the full three-phase protocol. The cost function decreases rapidly in Phase I. Its improvement slows in Phase II. It stops improving in Phase III. The gradient norm oscillates in Phase I. It then decays smoothly in Phase II. In Phase III, it becomes small. This indicates the onset of a BP.

Table~\ref{tab:checkpoints} shows that the cost function stops improving before the gradient norm vanishes completely. This means that optimization slows before the gradient disappears.

\begin{table}[htbp]
\centering
\caption{\footnotesize{Checkpoints during optimization.}}
\label{tab:checkpoints}
\begin{tabular}{lcccccr}
\hline
\hline
\textbf{Iter.} & \textbf{Cost} & $\bm{\eta}$ & $\bm{\mathcal{C}_{\text{rel}}}$ & $\bm{\mathcal{D}_{\text{rel}}}$ & \textbf{GradNorm} & \textbf{Ent} \\
\hline
0 & -0.7308 & 0.2344 & 0.9374 & 3.0612 & 2.628090 & 1.5684 \\
30 & -3.3106 & 0.1076 & 0.3938 & 3.2647 & 1.755816 & 1.1436 \\
60 & -4.7762 & 0.0192 & 0.0648 & 3.3163 & 0.938294 & 0.5977 \\
90 & -5.1018 & 0.0022 & 0.0074 & 3.3215 & 0.580100 & 0.7061 \\
120 & -5.1994 & 0.0002 & 0.0008 & 3.3219 & 0.215481 & 0.7521 \\
150 & -5.2166 & $0.000004$ & 0.0001 & 3.3219 & 0.0865 & 0.7640 \\
\hline
\end{tabular}
\end{table}

The CMCC $\Lambda_{\alpha,\beta}$ with $\beta=0.74$ converts coherence into intra-block mixedness. This reduces $\eta$. Adam's updates push parameters into low-energy regions, where coherence offers little advantage. This accelerates the loss of coherence.

Figure~\ref{figure3}(b) compares $\eta$ with entanglement entropy. As shown in Table~\ref{tab:analysis_summary}, $\eta$ decreases monotonically from about 0.2344 to roughly $0.000004$. This is a drop of about $99.998\%$. Entanglement entropy decreases slightly at first, then oscillates, and later becomes flat. This difference shows that $\eta$ is more sensitive to the conversion process than entanglement entropy.

\begin{table}[htbp]
\centering
\caption{\footnotesize{Summary of $\eta$ evolution and correlation between $\mathcal{C}_{\text{rel}}$ and $\mathcal{D}_{\text{rel}}$ at $\beta = 0.74$.}}
\label{tab:analysis_summary}
\begin{tabular}{lc}
\hline
\textbf{Parameter} & \textbf{Value} \\
\hline
$\beta$ & 0.74 \\
Initial $\eta$ & 0.2344 \\
Final $\eta$ & 0.000004 \\
Decrease in $\eta$ & 99.998\% \\
Correlation $\mathcal{C}_{\text{rel}}$ vs $\mathcal{D}_{\text{rel}}$ & $-0.9923$ \\
$p$-value & $5.99\times 10^{-107}$ \\
\hline
\end{tabular}
\end{table}

Figure~\ref{figure3}(c) compares the three metrics after rescaling them to $[0,1]$. The metric $\eta$ reaches the $24\%$ threshold at iteration 19. The gradient norm reaches it at iteration 30, and entanglement entropy at iteration 31 (see Table~\ref{tab:warning_points}).

\begin{table}[htbp]
\centering
\caption{Early warning points detected by various metrics when they decline by 24\% relative to their initial values.}
\label{tab:warning_points}
\setlength{\tabcolsep}{12pt}
\begin{tabular}{lr}
\hline
\textbf{Metric} & \textbf{Warning Iteration} \\
\hline
Coherence fraction $\eta$ & 19 \\
GradNorm & 30 \\
Ent & 31 \\
\hline
\end{tabular}
\end{table}

The absolute Pearson correlation coefficient $|r|$ measures how tightly two variables are linearly related. Here, it links a diagnostic metric $X$ to performance gains $Y$. Examples of $X$ include coherence fraction $\eta$, gradient norm, or entanglement entropy. Formally, it is defined as:
\begin{equation}
r_{XY} = \frac{\sum_{i=1}^{n}(X_i - \bar{X})(Y_i - \bar{Y})}{\sqrt{\sum_{i=1}^{n}(X_i - \bar{X})^2 \sum_{i=1}^{n}(Y_i - \bar{Y})^2}},
\end{equation}
where $X_i$ and $Y_i$ are paired observations, $\bar{X}$ and $\bar{Y}$ are their respective means, and $n$ is the number of observations. The absolute value $|r| \in [0,1]$ indicates the strength of the linear association, with higher values representing stronger predictive relationships.

Figure~\ref{figure3}(d)shows the correlation of each metric with future performance. The coherence fraction $\eta$ gives $|r| = 0.872$ and $R^2 = 0.760$. The gradient norm gives $|r| = 0.807$ and $R^2 = 0.650$. Entanglement entropy gives $|r| = 0.810$ and $R^2 = 0.656$.

As shown in Table~\ref{tab:correlations}, $\eta$ has the highest $|r|$ and $R^2$ among the three metrics.

\begin{table}[htbp]
\centering
\caption{\footnotesize{Pearson correlation $|r|$ and explained variance $R^2$ for each metric with future performance gains.}}
\label{tab:correlations}
\setlength{\tabcolsep}{12pt}
\begin{tabular}{lcr}
\hline
\textbf{Metric} & $\mathbf{|r|}$ & \textbf{Explained Variance ($R^{2}$)} \\
\hline
Coherence fraction $\eta$ & 0.872 & 0.760 \\
GradNorm & 0.807 & 0.650 \\
Entanglement entropy & 0.810 & 0.656 \\
\hline
\end{tabular}
\end{table}

\subsection{Early Warning Across Multiple Decline Thresholds}
\label{subsec:multi_threshold_beta}

We compare the three metrics across several decline thresholds and different conversion pathways (controlled by $\beta$).

\begin{table}[htbp]
\centering
\caption{Iteration number at which each metric reaches a given decline threshold.}
\label{tab:multiple_threshold_warnings}
\begin{tabular}{lccr}
\hline
\textbf{Decline Threshold} & \textbf{$\eta$} & \textbf{GradNorm} & \textbf{Ent} \\
\hline
15\%  & 15 & 24 & 27 \\
24\%  & 19 & 30 & 31 \\
33\% & 23 & 35 & 35 \\
42\%  & 27 & 41 & 39 \\
50\%  & 31 & 47 & 44 \\
60\%  & 35 & 60 & 149 \\
70\%  & 41 & 81 & -- \\
75\%  & 44 & 90 & -- \\
80\% & 48 & 99 & -- \\
90\%  & 59 & 118 & -- \\
\hline
\end{tabular}
\end{table}

\begin{figure}[ht]
\centering
\includegraphics[width=\linewidth]{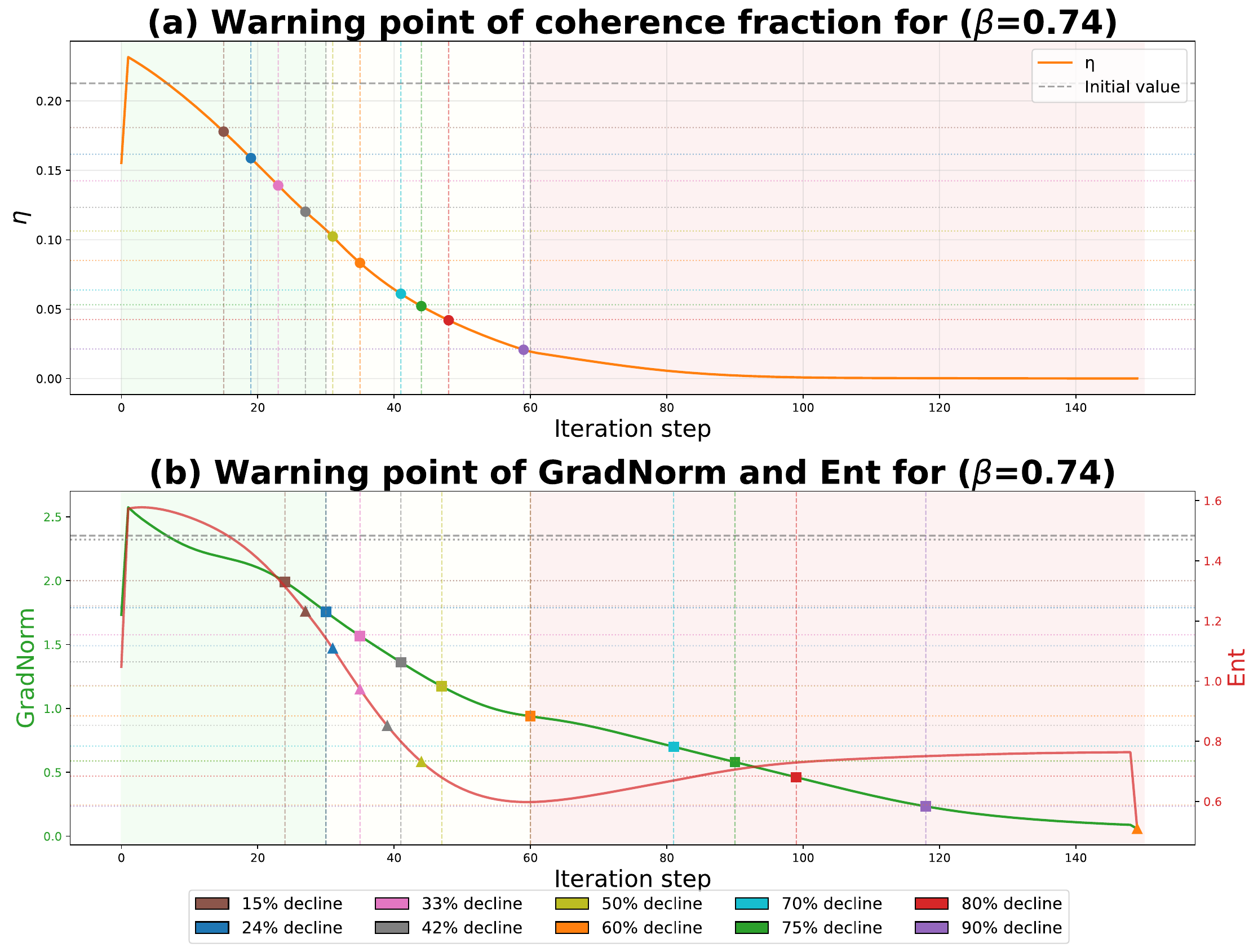}
\caption{\footnotesize{Iteration numbers at which each metric reaches selected decline thresholds.}}
\label{Figure3_multiple_threshold_warnings}
\end{figure}

As shown in Table~\ref{tab:multiple_threshold_warnings} and Fig.~\ref{Figure3_multiple_threshold_warnings}, $\eta$ reaches each threshold earlier than the gradient norm and entanglement entropy. For example, at the $15\%$ threshold, $\eta$ reaches it at iteration 15, the gradient norm at 24, and entanglement entropy at 27. At the $50\%$ threshold, $\eta$ reaches it at iteration 31, the gradient norm at 47, and entanglement entropy at 44. Entanglement entropy does not reach the $70\%$, $75\%$, $80\%$, or $90\%$ thresholds within 150 iterations.

\begin{table}[htbp]
\centering
\caption{Iteration numbers at which each metric reaches selected decline thresholds for different $\beta$ values.}
\label{tab:beta_sweep_warnings}
\begin{tabular}{lcccr}
\hline
\textbf{$\beta$} & \textbf{$\eta$ (15\%)} & \textbf{$\eta$ (24\%)} & \textbf{$\eta$ (50\%)} & \textbf{$\eta$ (75\%)}\\
\hline
0.1 & 39 & 44 & 54 & 64  \\
0.3 & 13 & 17 & 30 & 46  \\
0.5 & 13 & 36 & 46 & 57\\
0.7 & 12 & 14 & 21 & 30 \\
0.9 & 10 & 13 & 31 & 51  \\
\hline
\textbf{$\beta$} & \textbf{Grad (15\%)} & \textbf{Grad (24\%)} & \textbf{Grad (50\%)} & \textbf{Grad (75\%)} \\
\hline
0.1 & 66 & 72 & 85 & 107\\
0.3 & 22 & 36 & 58 & 100\\
0.5 & 24 & 32 & 56 & 82\\
0.7 & 17 & 21 & 33 & 115\\
0.9 & 14 & 22 & 54 & 85\\
\hline
\textbf{$\beta$} & \textbf{Ent (15\%)} & \textbf{Ent (24\%)} & \textbf{Ent (50\%)} & \textbf{Ent (75\%)} \\
\hline
0.1 & 59 & 60 & 64 & -- \\
0.3 & 20 & 31 & -- & -- \\
0.5 & 32 & 37 & 50 & 149 \\
0.7 & 74 & 79 & 94 & 149 \\
0.9 & 149 & 149 & -- & -- \\
\hline
\end{tabular}
\end{table}

\begin{figure}[ht]
\centering
\includegraphics[width=\linewidth]{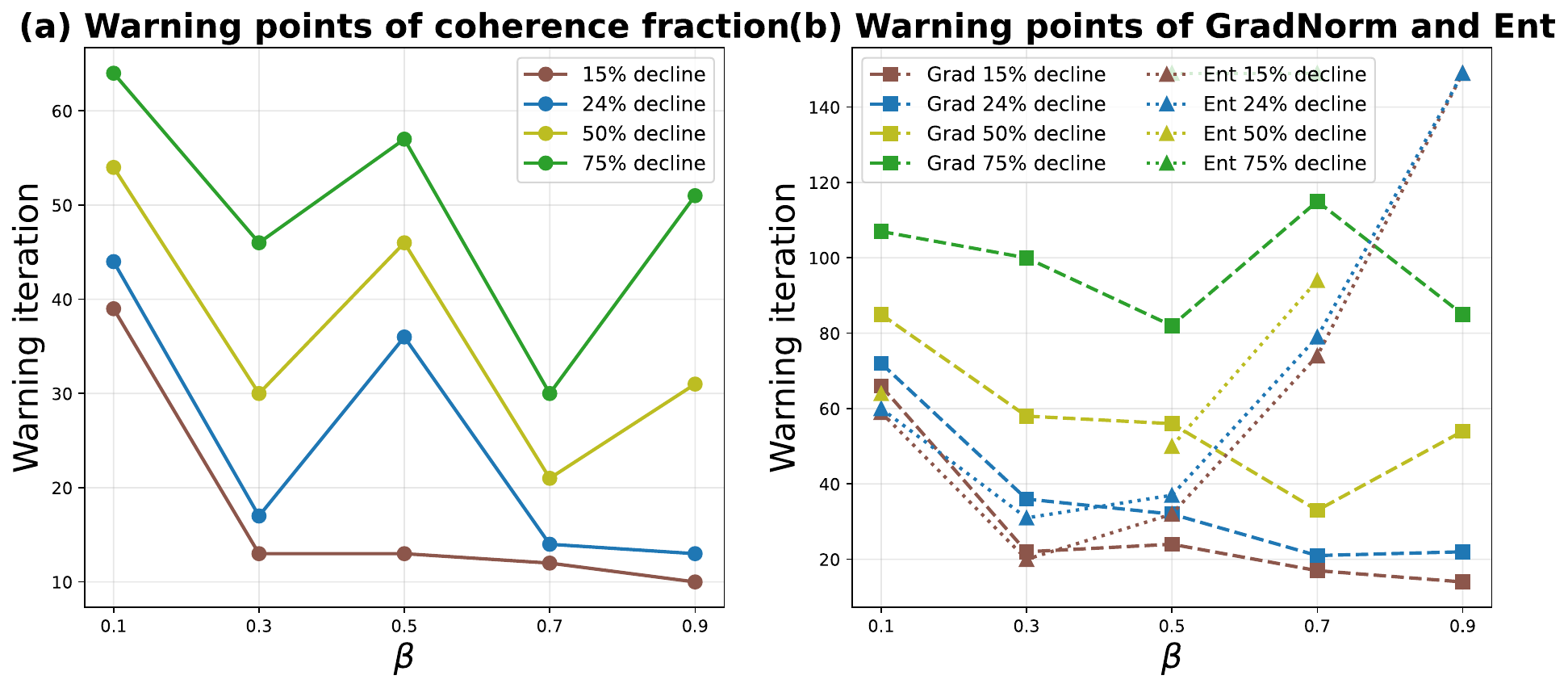}
\caption{\footnotesize{Iteration numbers at which each metric reaches selected decline thresholds for different $\beta$ values.}}
\label{Figure4_beta_sweep_warning_distribution}
\end{figure}

Figure~\ref{Figure4_beta_sweep_warning_distribution} and Table~\ref{tab:beta_sweep_warnings} show the results for $\beta$ values from 0.1 to 0.9. The metric $\eta$ reaches the thresholds earlier than the gradient norm and entanglement entropy for most $\beta$ values. For example, at $\beta = 0.1$, $\eta$ reaches the $15\%$ threshold at iteration 39. The gradient norm reaches it at iteration 66, and entanglement entropy at iteration 59. As $\beta$ increases, $\eta$ remains the earliest.

This behavior is not specific to a particular noise mixture. It reflects the conversion of coherence into mixedness. For larger $\beta$, entanglement entropy sometimes reaches the thresholds later, or not at all.

The metric $\eta$ captures the coherence-to-mixedness conversion that drives BPs. It can be monitored in real time. This allows adaptive adjustments of learning rates or parameter reinitialization before gradients vanish. Such monitoring is useful for near-term quantum hardware, where noise conditions vary.
\subsection{Summary and Discussion}
\label{subsec:comparative_summary}

Our comparison shows that $\eta$ has several advantages over the gradient norm and entanglement entropy as a diagnostic for VQA optimization.

\textbf{Early detection:} The metric $\eta$ reaches a given decline threshold earlier than the other two metrics. This holds across all tested thresholds and $\beta$ values. It often reaches the threshold several iterations before the gradient norm does.

\textbf{Correlation with performance:} The metric $\eta$ has a stronger linear correlation with future performance gains than the gradient norm or entanglement entropy.

\textbf{Mechanistic interpretation:} The gradient norm measures trainability. Entanglement entropy measures quantum correlations. The metric $\eta$ measures the conversion of coherence into mixedness. This connects optimization performance to the underlying resource dynamics. It also explains why optimization can stall even when gradients are not zero.

\textbf{Stability:} The behavior of $\eta$ is consistent across different conversion strengths and noise models. It does not depend on a specific setup.

\textbf{Practical use:} The metric $\eta$ can be computed during optimization. It can guide adjustments of the learning rate, circuit depth, or parameters. This provides a way to detect and respond to coherence loss before the gradient vanishes.

In summary, $\eta$ gives an earlier and more informative signal of coherence loss than standard diagnostics. It connects resource theory to optimization behavior and offers a practical tool for monitoring and improving VQA performance.

\section{Three-Dimensional Analysis of the Coherence Fraction}
\label{sec:3d_analysis}

Previous studies examined $\eta$ for fixed $\beta$ values. We extend the analysis to a full range of $\beta$.

\subsection{Global Behavior of coherence fraction}
\label{subsec:3d_method}

We vary $\beta$ over 30 values between 0 and 1. For each $\beta$, we run 150 optimization iterations. At selected checkpoints, we record four quantities: $\eta$, the cost $f(\vec{\theta})$, the gradient norm $\|\nabla f\|$, and the entanglement entropy.

\begin{figure}[ht]
    \centering
    \includegraphics[width=\linewidth]{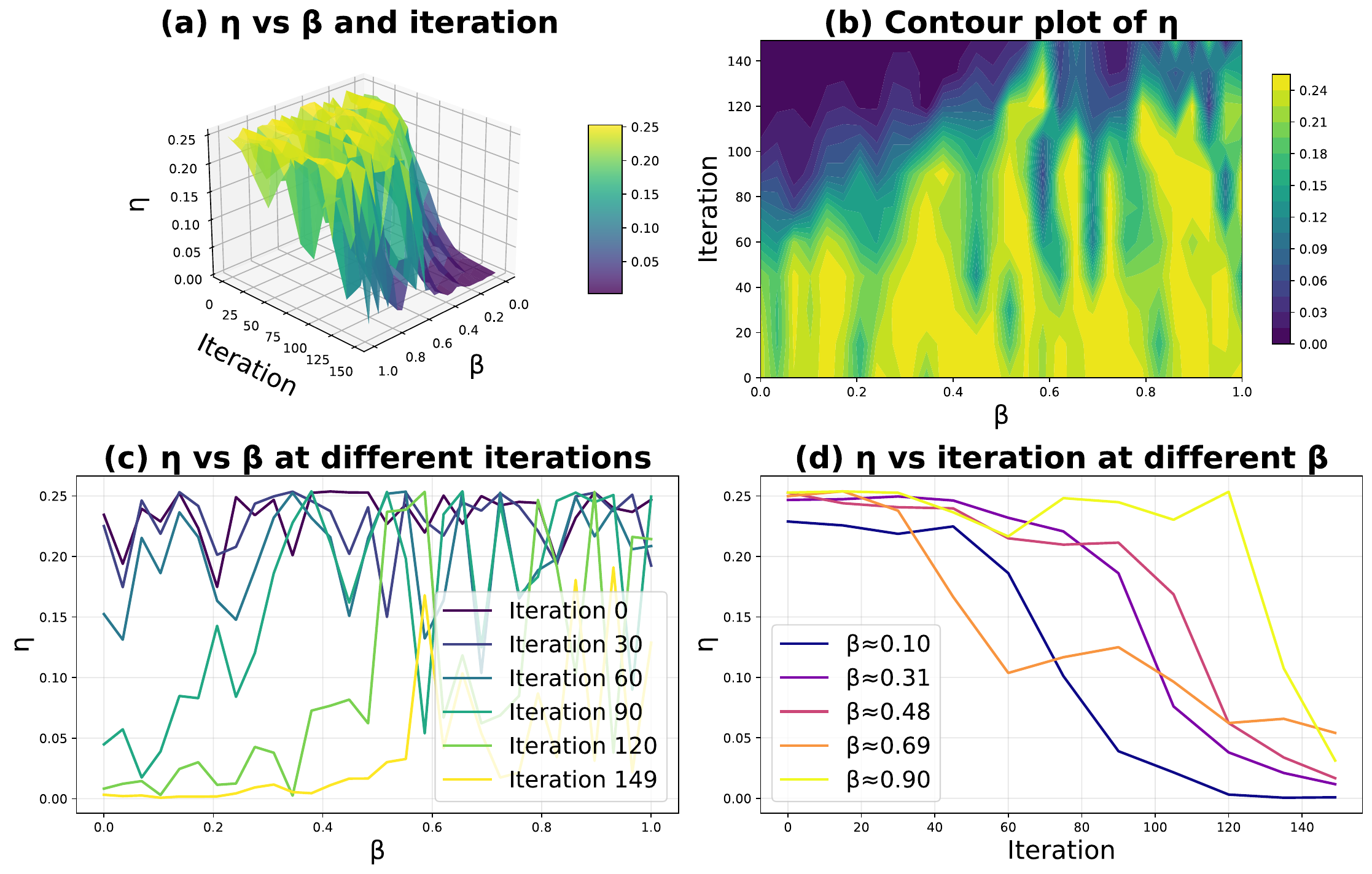}
    \caption{\footnotesize{Three-dimensional analysis of $\eta$ as a function of $\beta$ and iteration. (a) Surface plot of $\eta$ versus $\beta$ and iteration. (b) Contour map of $\eta$. (c) Cross-sections of the surface at fixed iterations (0, 30, 60, 90, 120, 149). (d) Cross-sections of the surface at fixed $\beta$ values (0.10, 0.31, 0.48, 0.69, 0.90). The value of $\eta$ decreases with iteration for all $\beta$, but the decay rate depends on $\beta$. For intermediate $\beta$ (e.g., 0.48), $\eta$ drops quickly at first and then stabilizes.}}
\label{fig:3d_eta_analysis}
\end{figure}

Figure~\ref{fig:3d_eta_analysis}(a) shows $\eta$ as a function of $\beta$ and iteration. The surface shows a non-monotonic dependence on $\beta$ and iteration. The effect of $\beta$ on $\eta$ depends on the optimization stage.

\begin{figure}[ht]
    \centering
    \includegraphics[width=\linewidth]{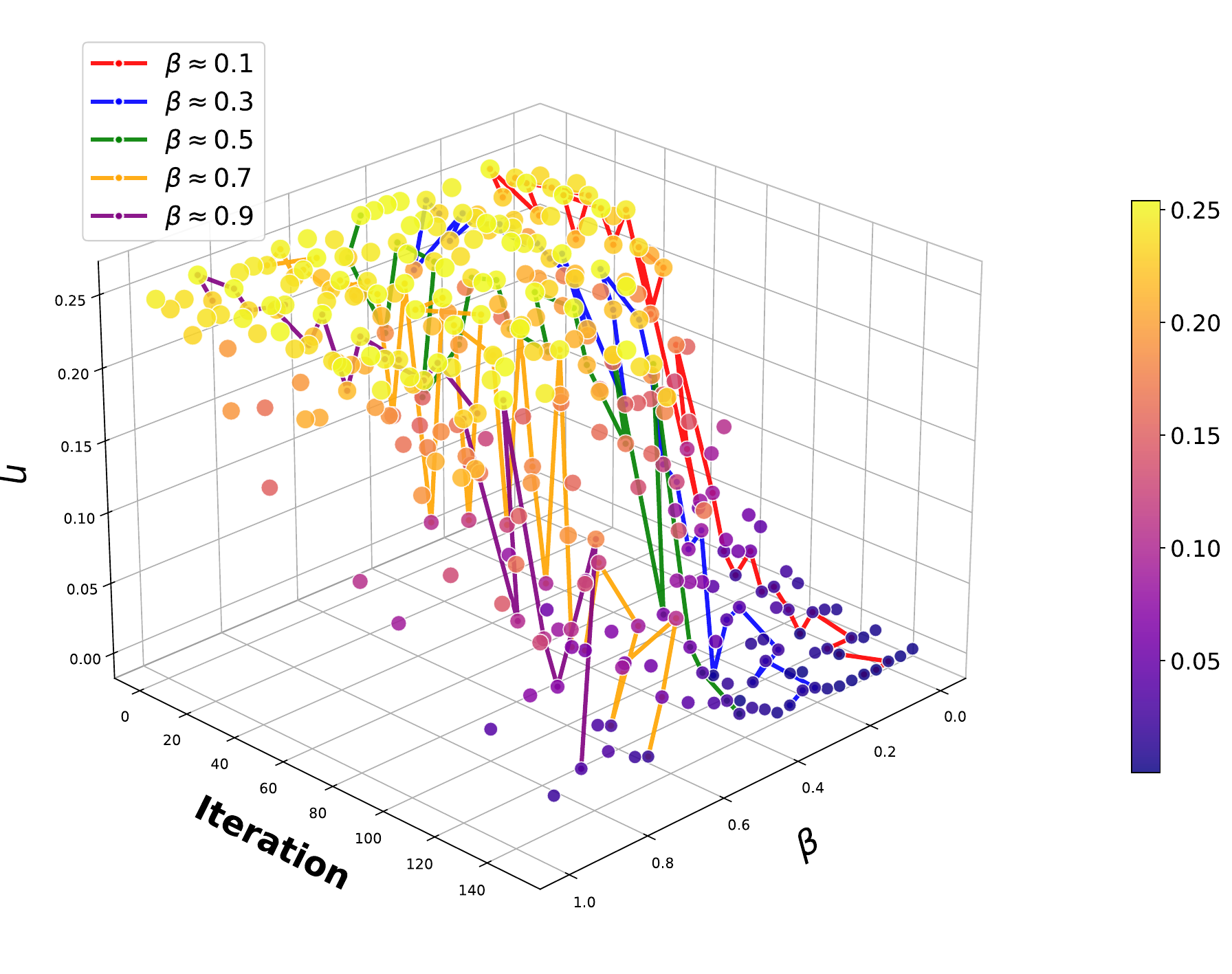}
    \caption{\footnotesize{Three-dimensional scatter plot of $\eta$ versus $\beta$ and iteration. Points are shown for selected $\beta$ values (about 0.1, 0.3, 0.5, 0.7, 0.9). The color and height of each point indicate the value of $\eta$.}}
    \label{fig:3d_scatter}
\end{figure}

At iteration 0, $\eta$ is similar for all $\beta$ values. At iteration 149, $\eta$ drops below $0.01$ for most $\beta$ values, and approaches zero in some regions. This indicates a loss of coherence.

The surface pattern shows that the effect of intra-block mixedness on $\eta$ depends on both $\beta$ and the optimization stage. In some regions, $\eta$ decreases rapidly as coherence is converted into mixedness.

The contour map in Figure~\ref{fig:3d_eta_analysis}(b) shows the $\eta$ surface from above. Darker regions correspond to higher $\eta$. These regions represent parameter choices where $\eta$ remains relatively high.

Figure~\ref{fig:3d_eta_analysis}(c) shows cross-sections of the surface at fixed iterations. Early iterations show little dependence on $\beta$. Later iterations show stronger dependence, with clear minima and maxima.

Figure~\ref{fig:3d_eta_analysis}(d) shows $\eta$ versus iteration for selected $\beta$ values. All curves start between $0.20$ and $0.25$. By iteration 150, they range from about $0.0001$ to $0.05$, depending on $\beta$.

The decay of $\eta$ depends on $\beta$. A sustained decrease in $\eta$ precedes the vanishing of gradients and the onset of BPs.

Figure~\ref{fig:3d_scatter} shows $\eta$ as a function of $\beta$ and iteration for selected $\beta$ values. The color and height of each point represent $\eta$. The trajectories show how $\eta$ changes with iteration for each $\beta$.

\subsection{Comparison with Gradient Norm and Entanglement Entropy}
\label{subsec:3d_comparative}

\begin{figure}[ht]
    \centering
    \includegraphics[width=\linewidth]{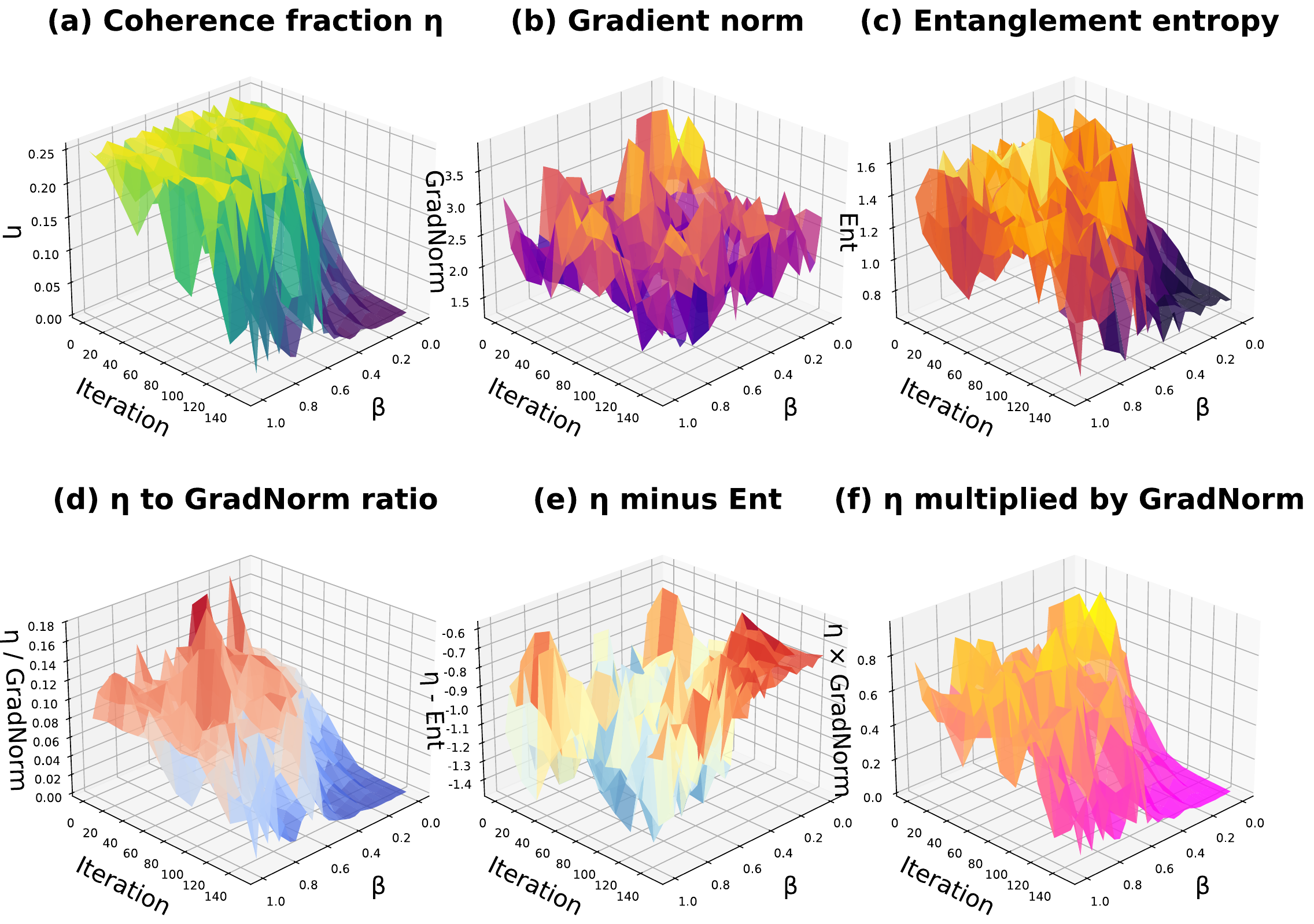}
    \caption{\footnotesize{Three-dimensional comparison of VQA metrics. (a) Coherence fraction $\eta$. (b) Gradient norm $\|\nabla f\|$. (c) Entanglement entropy. (d) Ratio $\eta / \|\nabla f\|$. (e) Difference $\eta - \text{Ent}$. (f) Product $\eta \times \|\nabla f\|$.}}
    \label{fig:3d_comparative}
\end{figure}

Figure~\ref{fig:3d_comparative} shows six quantities as functions of $\beta$ and iteration: $\eta$, the gradient norm, entanglement entropy, the ratio $\eta / \|\nabla f\|$, the difference $\eta - \text{Ent}$, and the product $\eta \times \|\nabla f\|$.

The surface of $\eta$ in (a) is non-monotonic. It depends on both $\beta$ and the optimization stage. Some regions of the $(\beta, \text{iteration})$ plane give higher $\eta$ than others.

The gradient norm surface in (b) decreases with $\beta$ and iteration. For $\beta > 0.8$, the gradient norm becomes very small by iteration 70. For $\beta < 0.2$, it remains above $10^{-1}$ throughout. The decrease of $\eta$ precedes the decrease of the gradient norm.

The entanglement entropy surface in (c) is non-monotonic. It has a maximum around $\beta \approx 0.5$ and iteration 50. The behavior of entanglement entropy is less sensitive to $\beta$ than $\eta$.

The ratio $\eta / \|\nabla f\|$ in (d) has a maximum (about 0.18) at $\beta \approx 0.32$ and iteration 58. For $\beta$ between 0.3 and 0.6, the ratio stays above 0.10 up to iteration 100. For $\beta > 0.7$, it drops below 0.05 by iteration 80.

The difference $\eta - \text{Ent}$ in (e) is negative throughout. The magnitude $|\eta - \text{Ent}|$ is large for $\beta > 0.6$ and iterations beyond 90, often exceeding 1.4.

The product $\eta \times \|\nabla f\|$ in (f) is largest for $\beta$ around 0.3 to 0.5 and iterations around 40 to 80. It becomes small for large $\beta$ and late iterations.

Together, these plots show that $\eta$ decreases as coherence is converted into mixedness. The decrease is largest for dephasing-dominated conversion (large $\beta$). The difference $\eta - \text{Ent}$ is always negative, with large magnitudes in the high-$\beta$, late-iteration region. Monitoring $\eta$ provides an earlier indication of optimization slowdown than the gradient norm.

\subsection{Cross-Sectional and Correlation Analysis}
\label{subsec:cross_correlation}

\begin{figure}[ht]
    \centering
    \includegraphics[width=\linewidth]{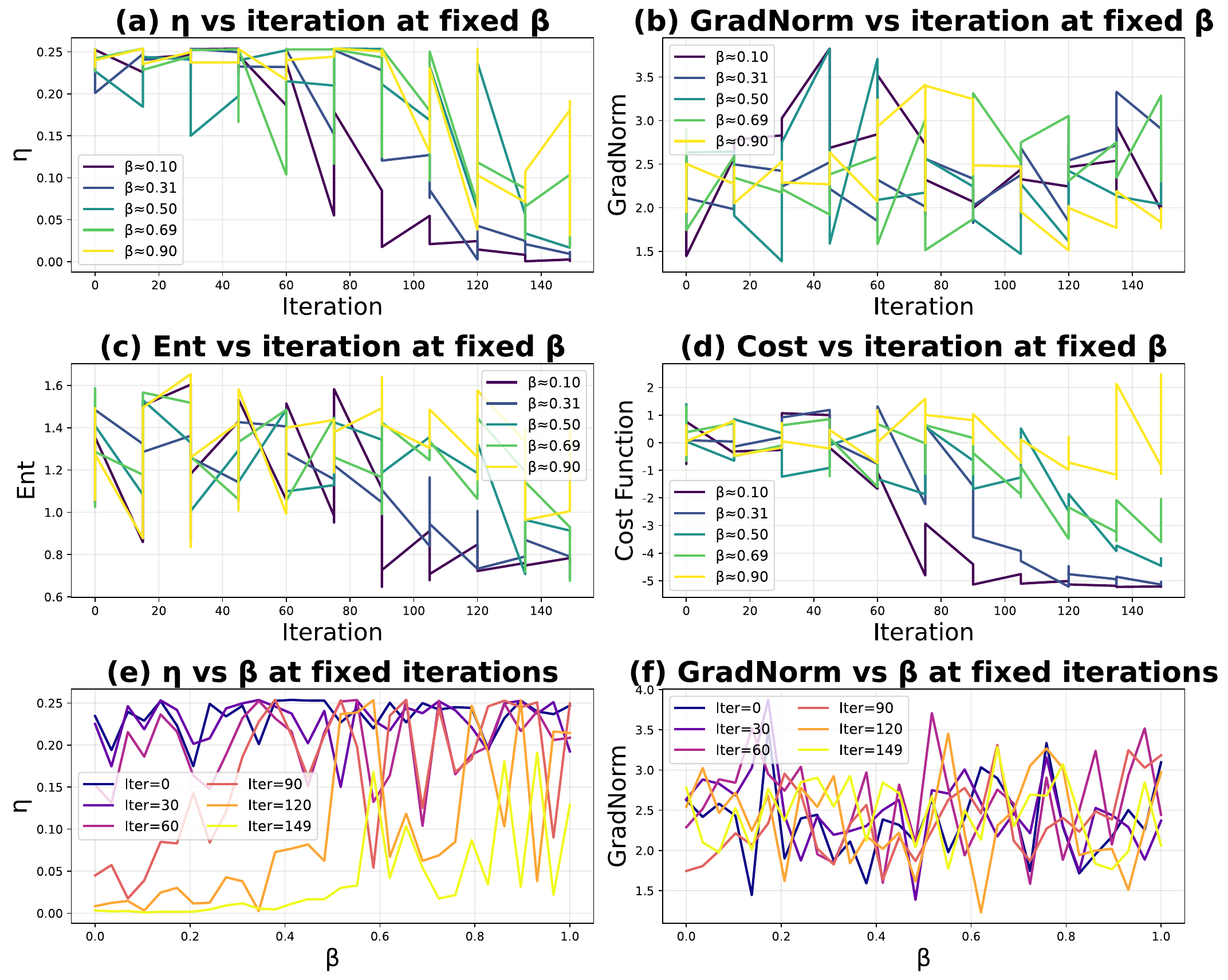}
    \caption{
    \footnotesize{Cross-sectional analysis of VQA metrics. (Top row) $\eta$, gradient norm, entanglement entropy, and cost versus iteration for selected fixed $\beta$ values. (Bottom row) $\eta$ and gradient norm versus $\beta$ at fixed iterations.}}
    \label{fig:cross_section}
\end{figure}

Figure~\ref{fig:cross_section} shows cross-sections of the metrics as functions of iteration for fixed $\beta$, and as functions of $\beta$ at fixed iterations.

As shown in Fig.~\ref{fig:cross_section}(a), $\eta$ decreases faster for larger $\beta$. The decrease is monotonic for all $\beta$ values. This reflects the conversion of block coherence $\mathcal{C}_{\text{rel}}$ into intra-block mixedness $\mathcal{D}_{\text{rel}}$.

Fig.~\ref{fig:cross_section}(b) shows the gradient norm versus iteration for the same $\beta$ values. The decrease in $\eta$ precedes the decrease in the gradient norm.

Fig.~\ref{fig:cross_section}(c) shows entanglement entropy versus iteration. Its behavior shows little dependence on $\beta$, in contrast to $\eta$.

Fig.~\ref{fig:cross_section}(d) shows the cost function versus iteration. The cost decreases fastest in early iterations, when $\eta$ is still relatively high. It then flattens as $\eta$ decreases.

Fig.~\ref{fig:cross_section}(e) shows $\eta$ versus $\beta$ at selected iterations. Fig.~\ref{fig:cross_section}(f) shows the gradient norm versus $\beta$ at the same iterations. Regions with low $\eta$ correspond to regions with small gradient norms.

The results show that $\eta$ decreases across all $\beta$ values and precedes the decrease of the gradient norm. The dependence of $\eta$ on $\beta$ indicates that the conversion of coherence into mixedness is a driver of optimization slowdown in VQAs.

We compute Pearson correlation coefficients across all sampled iterations and $\beta$ values from the 3D scan.

\begin{figure}[ht]
    \centering
    \includegraphics[width=\linewidth]{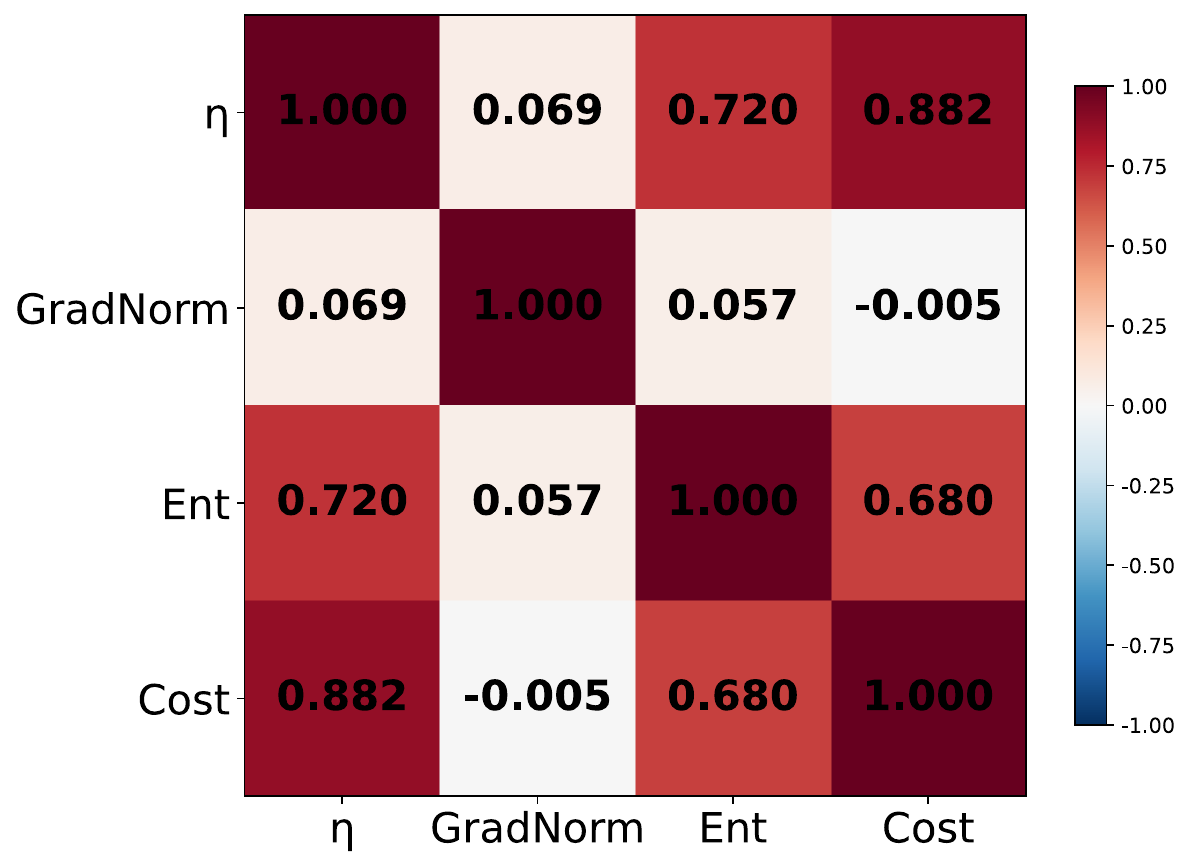}
    \caption{
    \footnotesize{Pearson correlation matrix of VQA metrics.}}
    \label{fig:cross_matrix}
\end{figure}

The correlation matrix shows the following patterns:
\begin{itemize}
    \item $\eta$ vs. Cost ($r = 0.882$): Higher $\eta$ is associated with lower cost.
    \item $\eta$ vs. Entanglement Entropy ($r = 0.720$): $\eta$ and entanglement entropy are positively correlated, but the correlation is moderate.
    \item $\eta$ vs. Gradient Norm ($r = 0.069$): $\eta$ and the gradient norm are nearly uncorrelated.
    \item Entanglement vs. Cost ($r = 0.680$): Entanglement entropy is positively correlated with cost, but less strongly than $\eta$.
    \item Gradient Norm vs. Cost ($r = -0.005$): The gradient norm is nearly uncorrelated with cost.
    \item Gradient Norm vs. Entanglement ($r = 0.057$): The gradient norm and entanglement entropy are nearly uncorrelated.
\end{itemize}

These correlations show that $\eta$ is more strongly associated with cost than the gradient norm or entanglement entropy.

\subsection{Summary and Discussion}

The 3D analysis shows that $\eta$ depends on both $\beta$ and the optimization stage. In some regions of the parameter space, $\eta$ remains relatively high. In others, it decreases rapidly.

The comparison with gradient norms and entanglement entropy shows that $\eta$ is a useful metric for monitoring optimization. These results provide a basis for real-time monitoring of quantum resources in NISQ devices.

\section{Discussion and Conclusions}
\label{con}

This work connects algorithmic performance to the conversion of coherence into intra-block mixedness. We track this conversion explicitly and show how it affects optimization on noisy devices. Our framework links resource theory to variational optimization and supports real-time monitoring of $\eta$.

The metric $\eta$ enables closed-loop control strategies, such as adaptive learning rate adjustment, dynamical decoupling, and parameter re-initialization. These strategies can reduce the loss of coherence and delay the onset of BPs. Experiments show that interventions guided by $\eta$ can improve optimization performance.

The conversion of coherence into intra-block mixedness is a neutral process. It becomes harmful only because current VQA architectures are designed to use coherence, not mixedness. A long-term solution is to develop algorithms that can use a broader range of resource forms. The metric $\eta$ helps identify this mismatch.

This work has several limitations. First, we use only a 4-qubit model. Future work could extend the analysis to larger systems. Second, we do not consider non-Markovian or device-specific noise. Third, computing $\eta$ in real time requires state estimation and entropy calculation, which may cause measurement overhead on NISQ devices. We have not discussed this issue in detail.

For future work, we plan to incorporate $\eta$ into optimizer design, for example in an Adam-$\eta$ variant. We also plan to explore adaptive coarse-graining strategies to control measurement overhead and improve optimization performance.
\section*{Acknowledgments}

The authors would like to thank \v{S}afr\'anek Dominik, Yunlong Xiao and Junjing Xing for their valuable discussions and assistance during this research.

\section*{Funding}

This research was funded by the Jiangxi Province Key Laboratory of Applied Optical Technology [2024SSY03051].

\section*{DATA AVAILABILITY}

All data that support the findings of this study are included within the article (and any supplementary files).

\section*{Competing Interests}

All authors declare no financial or non-financial competing interests.

\section*{Author Contributions}

X.Z. conceived the research, developed the theoretical framework, designed and performed all numerical simulations, analyzed the data, and wrote the manuscript. X.Z. also served as the corresponding author.

\begin{appendices}

\section{Proof of Proposition~\ref{prop:eta_properties}}\label{APPxA}

\begin{proof}
1. \textbf{Boundedness:} Since $\mathcal{C}_{\text{rel}}(\rho, \mathcal{C}) \geq 0$ and $\mathcal{D}_{\text{rel}}(\rho) \geq 0$, we have
\begin{align}
\mathcal{O}_{\mathcal{C}}(\rho) = \mathcal{C}_{\text{rel}}(\rho, \mathcal{C}) + \mathcal{D}_{\text{rel}}(\rho) \geq \mathcal{C}_{\text{rel}}(\rho, \mathcal{C}).
\end{align}
  
Therefore,
\begin{align}
\eta(\rho) = \mathcal{C}_{\text{rel}}(\rho, \mathcal{C}) / \mathcal{O}_{\mathcal{C}}(\rho) \leq 1.
\end{align}
   
The lower bound $\eta(\rho) \geq 0$ follows from the non-negativity of the numerator and denominator. By definition, $\eta(\rho)=0$ when $\mathcal{O}_{\mathcal{C}}(\rho)=0$.

2. \textbf{Monotonicity:} For any free operation $\Lambda$, the coherence measure $\mathcal{C}_{\text{rel}}$ is non-increasing:
\begin{align}
\mathcal{C}_{\text{rel}}(\Lambda(\rho), \mathcal{C}) \leq \mathcal{C}_{\text{rel}}(\rho, \mathcal{C}).
\end{align}
  
The coherence capacity $\mathcal{O}_{\mathcal{C}}$ is not necessarily monotonic under free operations. Define
\begin{align}
\begin{aligned}
\Delta\mathcal{C}_{\text{rel}} &= \mathcal{C}_{\text{rel}}(\Lambda(\rho)) - \mathcal{C}_{\text{rel}}(\rho) \leq 0,\\
\Delta\mathcal{D}_{\text{rel}} &= \mathcal{D}_{\text{rel}}(\Lambda(\rho)) - \mathcal{D}_{\text{rel}}(\rho), \\ 
\Delta\mathcal{O}_{\mathcal{C}} &= \Delta\mathcal{C}_{\text{rel}} + \Delta\mathcal{D}_{\text{rel}}.
\end{aligned}
\end{align}

Then
 \begin{align}
\eta(\Lambda(\rho)) - \eta(\rho) = \frac{\mathcal{C}_{\text{rel}}(\rho) + \Delta\mathcal{C}_{\text{rel}}}{\mathcal{O}_{\mathcal{C}}(\rho) + \Delta\mathcal{O}_{\mathcal{C}}} - \frac{\mathcal{C}_{\text{rel}}(\rho)}{\mathcal{O}_{\mathcal{C}}(\rho)}.
\end{align}

The denominator is positive. After cross-multiplication, the sign of the difference is determined by
\begin{align}
\begin{aligned}
 &\big( \mathcal{C}_{\text{rel}}(\rho) + \Delta\mathcal{C}_{\text{rel}} \big) \, \mathcal{O}_{\mathcal{C}}(\rho) - \mathcal{C}_{\text{rel}}(\rho) \, \big( \mathcal{O}_{\mathcal{C}}(\rho) + \Delta\mathcal{O}_{\mathcal{C}} \big) \\ 
&\quad= \Delta\mathcal{C}_{\text{rel}} \, \mathcal{O}_{\mathcal{C}}(\rho) - \mathcal{C}_{\text{rel}}(\rho) \, \Delta\mathcal{O}_{\mathcal{C}}.
\end{aligned}
\end{align}

Since $\Delta\mathcal{C}_{\text{rel}} \leq 0$ and $\mathcal{O}_{\mathcal{C}}(\rho) \geq \mathcal{C}_{\text{rel}}(\rho)$, the expression is non-positive whenever $\Delta\mathcal{O}_{\mathcal{C}} \geq 0$. The case $\Delta\mathcal{O}_{\mathcal{C}} < 0$ requires closer examination.

As shown in this work, free operations can reduce coherence, but they cannot convert intra-block mixedness into coherence. From the data-processing inequality for $\mathcal{C}_{\text{rel}}$ and the decomposition, no free operation can increase $\eta$.

Suppose, for contradiction, that there exists a free operation $\Lambda$ such that $\eta(\Lambda(\rho)) > \eta(\rho)$. Then the relative loss in $\mathcal{C}_{\text{rel}}$ would be smaller than the relative loss in $\mathcal{O}_{\mathcal{C}}$. This would mean that the operation converts intra-block mixedness into coherence.

But this is impossible: free operations cannot generate coherence. Therefore,
\begin{align}
\eta(\Lambda(\rho)) \leq \eta(\rho),
\end{align}
for every free operation $\Lambda$.

3. \textbf{Strict decrease under CMCC:} Consider the channel $\Lambda_{\alpha,\beta}$. With probability $\alpha$, it replaces the input with a block-diagonal state (either fully dephased or projected). This removes coherence: it reduces $\mathcal{C}_{\text{rel}}$ unless the input is already block-incoherent.

From Theorem~\ref{Thm1}, $\mathcal{O}_{\mathcal{C}}$ is approximately conserved for suitable parameter choices.

Thus the numerator of $\eta$ (which depends on $\mathcal{C}_{\text{rel}}$) decreases, while the denominator ($\mathcal{O}_{\mathcal{C}}$) stays nearly constant. Hence $\eta$ decreases strictly under $\Lambda_{\alpha,\beta}$.
\end{proof}

\section{Proof of Theorem~\ref{Thm1}}\label{APPxB}

\begin{proof}
\textbf{Step 1: System and coarse-graining setup}

Consider a four-dimensional system with Hilbert space $\mathcal{H} = \mathbb{C}^4$ and computational basis 
\begin{align}
    \{|00\rangle, |01\rangle, |10\rangle, |11\rangle\}.
\end{align}

Define a coarse-graining $\mathcal{C} = \{\hat{P}_1, \hat{P}_2\}$ with
\begin{align}
\begin{aligned}
\hat{P}_1 &= |00\rangle\langle 00| + |01\rangle\langle 01|,\\
\hat{P}_2 &= |10\rangle\langle 10| + |11\rangle\langle 11|.
\end{aligned}
\end{align}

Each subspace has dimension $V_1 = V_2 = 2$. The maximally mixed reference states are
\begin{align}
\kappa_1 = \tfrac{1}{2}|00\rangle\langle 00| + \tfrac{1}{2}|01\rangle\langle 01|, 
\end{align}
\begin{align}
\kappa_2 = \tfrac{1}{2}|10\rangle\langle 10| + \tfrac{1}{2}|11\rangle\langle 11|.
\end{align}

\textbf{Step 2: Initial state}

Choose the initial state
\begin{align}
\begin{aligned}
\rho & = \tfrac{3}{8}|00\rangle\langle 00| + \tfrac{1}{4}|00\rangle\langle 11| + \tfrac{1}{8}|01\rangle\langle 01| \\
&+ \tfrac{1}{8}|10\rangle\langle 10| + \tfrac{1}{4}|11\rangle\langle 00| + \tfrac{3}{8}|11\rangle\langle 11| .
\end{aligned}
\end{align}

This state has off-diagonal terms between the two subspaces, so $\mathcal{C}_{\text{rel}}(\rho,\mathcal{C}) > 0$. Its eigenvalues are $\lambda_1 = 5/8$ and $\lambda_2 = \lambda_3 = \lambda_4 = 1/8$.

\textbf{Step 3: Baseline resource metrics}

\textbf{(a) $\mathcal{C}_{\text{rel}}(\rho,\mathcal{C})$}

The block-diagonal part of $\rho$ is
\begin{align}
\begin{aligned}
\hat{P}_{1}\rho\hat{P}_{1}&=\frac{3}{8}|00\rangle\langle 00| +\frac{1}{8}|01\rangle\langle 01|,\\
\hat{P}_{2}\rho\hat{P}_{2}&=\frac{1}{8}|10\rangle\langle 10| +\frac{3}{8}|11\rangle\langle 11|.
\end{aligned}
\end{align}

Using $\mathcal{C}_{\text{rel}}(\rho, \mathcal{C}) = S(\Delta(\rho)) - S(\rho)$, we have
\begin{align}
\Delta(\rho) = \frac{3}{8}|00\rangle\langle 00|+\frac{1}{8}|01\rangle\langle 01|+\frac{1}{8}|10\rangle\langle 10|+\frac{3}{8}|11\rangle\langle 11|.
\end{align}

The entropies are
\begin{align}
\begin{aligned}
S(\rho)&= - \frac{5}{8}\log_{2}\frac{5}{8} - \frac{3}{8}\log_{2}\frac{3}{8},\\
S(\Delta(\rho))&= - \frac{3}{4}\log_{2}\frac{3}{8} - \frac{1}{4}\log_{2}\frac{1}{8}.
\end{aligned}
\end{align}

Thus
\begin{align}
\mathcal{C}_{\text{rel}}(\rho, \mathcal{C}) =  - \frac{3}{4}\log_{2}\frac{3}{8}+\frac{5}{8}\log_{2}\frac{5}{8}+\frac{1}{8}\log_{2}\frac{1}{8}.
\end{align}

\textbf{(b) $\mathcal{D}_{\text{rel}}(\rho)$}

The conditional states are
\begin{align}
\begin{aligned}
\rho_{1}&=\frac{3}{4}|00\rangle\langle 00| +\frac{1}{4}|01\rangle\langle 01|,\\
\rho_{2}&=\frac{1}{4}|10\rangle\langle 10| +\frac{3}{4}|11\rangle\langle 11|.
\end{aligned}
\end{align}

Their relative entropies with respect to $\kappa_1$ and $\kappa_2$ are
\begin{align}
\begin{aligned}
D[\rho_1 \Vert  \kappa_1] &= \frac{3}{4}\log_{2}3 - 1,\\
D[\rho_2 \Vert  \kappa_2] &= \frac{3}{4}\log_{2}3 - 1.
\end{aligned}
\end{align}

Since $p_1 = p_2 = 1/2$,
\begin{align}
\mathcal{D}_{\text{rel}}(\rho)=\frac{3}{4}\log_{2}3 - 1.
\end{align}

\textbf{(c) Coherence capacity}
\begin{align}
\begin{aligned}
\mathcal{O}_{\mathcal{C}}(\rho) 
&= \mathcal{C}_{\text{rel}}(\rho, \mathcal{C}) + \mathcal{D}_{\text{rel}}(\rho)\\
&= - \frac{3}{4}\log_{2}\frac{3}{8}+\frac{5}{8}\log_{2}\frac{5}{8}
+\frac{1}{8}\log_{2}\frac{1}{8}+\frac{3}{4}\log_{2}3 - 1.
\end{aligned}
\end{align}

\textbf{Step 4: The conversion channel}

Define
\begin{align}\label{D1}
\Lambda_\varepsilon(\rho) = (1 - \varepsilon)\rho + \varepsilon\,|00 \rangle\langle 00 |,
\end{align}
with $0 \leq \varepsilon \leq 1$.

This channel is free because it outputs a block-incoherent state $|00\rangle\langle 00|$ with probability $\varepsilon$. It removes coherence between blocks and increases intra-block mixedness. By tuning $\varepsilon$, we control the conversion strength.

Applied to $\rho$, the channel gives
\begin{align}
\begin{aligned}
\Lambda_\varepsilon(\rho) ={}& \frac{3+5\varepsilon}{8}\,|00\rangle\langle 00| + \frac{1 - \varepsilon}{4}\,|00\rangle\langle 11| \\
&+ \frac{1 - \varepsilon}{8}\,|01\rangle\langle 01| + \frac{1 - \varepsilon}{8}\,|10\rangle\langle 10| \\
&+ \frac{1 - \varepsilon}{4}\,|11\rangle\langle 00| + \frac{3 - 3\varepsilon}{8}\,|11\rangle\langle 11|.
\end{aligned}
\end{align}

Its eigenvalues are
\begin{align}
\lambda^{\varepsilon}_{1} = \lambda^{\varepsilon}_{2} = \frac{1 - \varepsilon}{8},
\end{align}
\begin{align}
\lambda^{\varepsilon}_{3} = \frac{3+\varepsilon - 2\sqrt{5\varepsilon^{2} - 2\varepsilon+1}}{8},
\end{align}
\begin{align}
\lambda^{\varepsilon}_{4} = \frac{3+\varepsilon + 2\sqrt{5\varepsilon^{2} - 2\varepsilon+1}}{8}.
\end{align}

\textbf{Step 5: Resource metrics after the channel}

\textbf{(a) $\mathcal{C}_{\text{rel}}(\Lambda_\varepsilon(\rho), \mathcal{C})$}

The block projections are
\begin{align}
\begin{aligned}
\hat{P}_1\Lambda_\varepsilon(\rho)\hat{P}_1 &= \frac{3+5\varepsilon}{8}\,|00\rangle\langle 00| + \frac{1 - \varepsilon}{8}\,|01\rangle\langle 01|,\\
\hat{P}_2\Lambda_\varepsilon(\rho)\hat{P}_2 &= \frac{1 - \varepsilon}{8}\,|10\rangle\langle 10| + \frac{3 - 3\varepsilon}{8}\,|11\rangle\langle 11|.
\end{aligned}
\end{align}

The dephased state is
\begin{align}
\begin{aligned}
\Delta(\Lambda_\varepsilon(\rho)) ={}& \frac{3+5\varepsilon}{8}\,|00\rangle\langle 00| + \frac{1 - \varepsilon}{8}\,|01\rangle\langle 01| \\
&+ \frac{1 - \varepsilon}{8}\,|10\rangle\langle 10| + \frac{3 - 3\varepsilon}{8}\,|11\rangle\langle 11|.
\end{aligned}
\end{align}

The entropies are
\begin{align}
\begin{aligned}
S(\Lambda_\varepsilon(\rho)) =& - \frac{1 - \varepsilon}{4}\log_{2}\frac{1 - \varepsilon}{8}
 - \lambda^{\varepsilon}_{3}\log_{2}\lambda^{\varepsilon}_{3}
 - \lambda^{\varepsilon}_{4}\log_{2}\lambda^{\varepsilon}_{4},
\end{aligned}
\end{align}
\begin{align}
\begin{aligned}
S(\Delta(\Lambda_\varepsilon(\rho))) =& - \frac{3+5\varepsilon}{8}\log_{2}\frac{3+5\varepsilon}{8}\\
& - \frac{1 - \varepsilon}{4}\log_{2}\frac{1 - \varepsilon}{8}\\
& - \frac{3 - 3\varepsilon}{8}\log_{2}\frac{3 - 3\varepsilon}{8}.
\end{aligned}
\end{align}

Thus
\begin{align}
\begin{aligned}
\mathcal{C}_{\text{rel}}(\Lambda_\varepsilon(\rho), \mathcal{C}) =& - \frac{3 - 3\varepsilon}{8}\log_{2}\frac{3 - 3\varepsilon}{8}\\
& - \frac{3+5\varepsilon}{8}\log_{2}\frac{3+5\varepsilon}{8} \\
&+ \lambda^{\varepsilon}_{3}\log_{2}\lambda^{\varepsilon}_{3} + \lambda^{\varepsilon}_{4}\log_{2}\lambda^{\varepsilon}_{4}.
\end{aligned}
\end{align}

\textbf{(b) $\mathcal{D}_{\text{rel}}(\Lambda_\varepsilon(\rho))$}

For $p^{\varepsilon}_1 =(1+\varepsilon)/2$ and $p^{\varepsilon}_2 = (1-\varepsilon)/2$,
\begin{align}
\begin{aligned}
\rho^{\varepsilon}_1 &= \frac{3+5\varepsilon}{4(1+\varepsilon)}\,|00\rangle\langle 00|+ \frac{1 - \varepsilon}{4(1+\varepsilon)}\,|01\rangle\langle 01|,\\
\rho^{\varepsilon}_2 &= \frac{1}{4}\,|10\rangle\langle 10| + \frac{3}{4}\,|11\rangle\langle 11|.
\end{aligned}
\end{align}

The relative entropies are
\begin{align}
\begin{aligned}
D[\rho^{\varepsilon}_1 \Vert \kappa_1] &= \frac{3+5\varepsilon}{4(1+\varepsilon)}\log_{2}\frac{3+5\varepsilon}{4(1+\varepsilon)}\\
 &+ \frac{1 - \varepsilon}{4(1+\varepsilon)}\log_{2}\frac{1 - \varepsilon}{4(1+\varepsilon)} + 1,\\
D[\rho^{\varepsilon}_2 \Vert \kappa_2] &= \frac{3}{4}\log_{2}3 - 1.
\end{aligned}
\end{align}

Therefore,
\begin{align}
\begin{aligned}
\mathcal{D}_{\text{rel}}(\Lambda_\varepsilon(\rho)) ={}& \frac{3+5\varepsilon}{8}\log_{2}\frac{3+5\varepsilon}{4(1+\varepsilon)}\\
 &+ \frac{1 - \varepsilon}{8}\log_{2}\frac{27(1 - \varepsilon)}{4(1+\varepsilon)} + \varepsilon.
\end{aligned}
\end{align}

\textbf{(c) Coherence capat}

\begin{align}
\begin{aligned}
\mathcal{O}_{\mathcal{C}}(\Lambda_\varepsilon(\rho)) ={}& \mathcal{C}_{\text{rel}}(\Lambda_\varepsilon(\rho), \mathcal{C}) + \mathcal{D}_{\text{rel}}(\Lambda_\varepsilon(\rho)) \\
={}& \lambda^{\varepsilon}_{3}\log_{2}\lambda^{\varepsilon}_{3}
 - \frac{3+5\varepsilon}{8}\log_{2}\frac{1+\varepsilon}{2}\\
 &- \frac{3 - 3\varepsilon}{8}\log_{2}\frac{1 - \varepsilon}{8} + \lambda^{\varepsilon}_{4}\log_{2}\lambda^{\varepsilon}_{4}\\
& + \frac{1 - \varepsilon}{8}\log_{2}\frac{1 - \varepsilon}{4(1+\varepsilon)} + \varepsilon.
\end{aligned}
\end{align}

\textbf{Numerical results at $\varepsilon = 0.26$}

\begin{align}
\begin{aligned}
\Delta \mathcal{C}_{\text{rel}}&= -0.12986, \\
\Delta \mathcal{D}_{\text{rel}} &= 0.12898,\\
\Delta \mathcal{O}_{\mathcal{C}} &= -0.00089,
\end{aligned}
\end{align}
and
\begin{align}
\begin{aligned}
\eta(\rho)&\approx 0.58174,\quad \eta(\Lambda_{\varepsilon}(\rho))\approx 0.29450,\\
\mathcal{O}_\mathcal{C}(\rho)&\approx 0.45121, \quad \frac{|\Delta \mathcal{C}_{\text{rel}}|}{\mathcal{O}_\mathcal{C}(\rho)}\approx 0.28735.
\end{aligned}
\end{align}

These results satisfy
\begin{align}
\begin{aligned}
&\Delta \mathcal{C}_{\text{rel}}< 0,\quad \Delta \mathcal{D}_{\text{rel}} > 0, \quad |\Delta \mathcal{D}_{\text{rel}}| \approx |\Delta \mathcal{C}_{\text{rel}}|,\\
&\Delta \mathcal{O}_{\mathcal{C}} \approx 0,\quad \eta(\Lambda_{\varepsilon}(\rho)) \approx \eta(\rho) - \frac{|\Delta \mathcal{C}_{\text{rel}}|}{\mathcal{O}_\mathcal{C}(\rho)}.
\end{aligned}
\end{align}

We find $\Delta\eta_{\%} \approx 49.376\%$. Thus $\eta$ decreases while $\mathcal{O}_{\mathcal{C}}$ remains approximately constant. This confirms the mechanism described in Theorem~\ref{Thm1}.
\end{proof}




\end{appendices}


\end{document}